\documentclass[ reprint, amssymb, amsmath, prb, showpacs,showkeys]{revtex4-1}
\usepackage{graphicx}
\usepackage{bm}

\usepackage[colorlinks=true,linkcolor=blue]{hyperref}
\expandafter\ifx\csname package@font\endcsname\relax\else
 \expandafter\expandafter
 \expandafter\usepackage
 \expandafter\expandafter
 \expandafter{\csname package@font\endcsname}
\fi
\usepackage{color}

\begin{document}

\title{Electronic band structure and ambipolar electrical properties of Cu$_2$O based semiconductor alloys}

\author{Vladan Stevanovi\'c}%
\affiliation{Colorado School of Mines, Golden, Colorado 80401, USA}%
\affiliation{National Renewable Energy Laboratory, Golden, Colorado 80401, USA}%

\author{Andriy Zakutayev}
\affiliation{National Renewable Energy Laboratory, Golden, Colorado 80401, USA}

\author{Stephan Lany}
\email{Stephan.Lany@nrel.gov}
\affiliation{National Renewable Energy Laboratory, Golden, Colorado 80401, USA}

\date{\today}%

\begin{abstract}
Tuning the opto-electronic properties through alloying is essential for semiconductor technology. 
Currently, mostly isovalent and isostructural alloys are used  (e.g., group-IV and III-V), but a vast and unexplored 
space of novel functional materials is conceivable when considering more complex alloys by mixing aliovalent and 
heterostructural constituents. The real challenge lies in the quantitative property prediction for such complex alloys to 
guide their experimental exploration. We developed an approach to predict compositional dependence of both 
band-structure and electrical properties from ab-initio calculations by extending conventional dilute defect model 
to higher (alloy) concentrations. Considering alloying of aliovalent (Mg, Zn, Cd) cations and isovalent anions (S, Se) into 
Cu$_2$O, we predict tunability of band-gap energies and doping levels over a wide range, including the type 
conversion from p- to n-type. Initial synthesis and characterization of Zn and Se substituted Cu$_2$O support 
the defect model, suggesting these alloys as promising novel oxide semiconductor materials.
\end{abstract}

\maketitle

\section*{Introduction}
%
Semiconductor alloys are typically mixtures of two isovalent and isostructural 
materials, e.g. Si$_{1-x}$Ge$_x$ in microelectronics \cite{brunner_RPP:2002,mayeul_PRL:2012}, 
Ga$_{1-x}$In$_x$N for blue light-emitting diodes \cite{nakamura_JJAP:1995}, or Cd$_{1-x}$Zn$_x$Te 
for radiation detectors \cite{knoll:1999}. In photovoltaics, 
the solar cells with the highest conversion efficiencies above 40 \% are multijunction devices with many 
layers of carefully engineered III-V alloys grown on a Ge substrate \cite{kinsey:2010}.
Whereas isovalent alloying typically employs compositions from a few per cent up to equal amounts of the 
constituents, so to modify the band-structure and optical properties, non-isovalent impurity doping \cite{woodyard_USP:1950}
is used to tailor the electrical properties via more dilute substitutions ranging from parts per million ($\sim$10$^{16}$cm$^{-3}$) 
up to few per cent in transparent conducting oxides \cite{ginley:2010}.

Accordingly, standard theoretical approaches of electronic structure calculations usually 
address either the modification of band-structure properties due to alloying \cite{lindsay_SSC:1999, kent_PRB:2001,popescu_PRL:2010} 
or the manipulation of electrical properties due to doping \cite{northrup_PRB:1993, lany_APL:2010,varley_PRB:2010}. 
However, a more general approach to semiconductor alloys includes the possibility of 
mixing aliovalent and heterostructural materials. In this case, the variation of band-structure and electrical 
properties is inherently coupled, and methods for describing alloy formation enthalpies need to include 
the Fermi energy as an additional variable that affects the formation enthalpy of non-isovalent substituents 
and eventually determines the carrier (electron or hole) concentrations in the alloy. A notable previous 
work in this regard is the study of Sm-doped CeO$_2$ by van de Walle and Ellis \cite{awalle_PRL:2007}, 
where the valence mismatch between Ce$^{\mathrm{+IV}}$ and Sm$^{\mathrm{+III}}$ is accommodated by formation of charged 
oxygen vacancy defects (V$_\mathrm{O}^{2+}$), but without the generation of free carriers. 

\begin{figure}[h!]
\centering
\includegraphics[width=\linewidth]{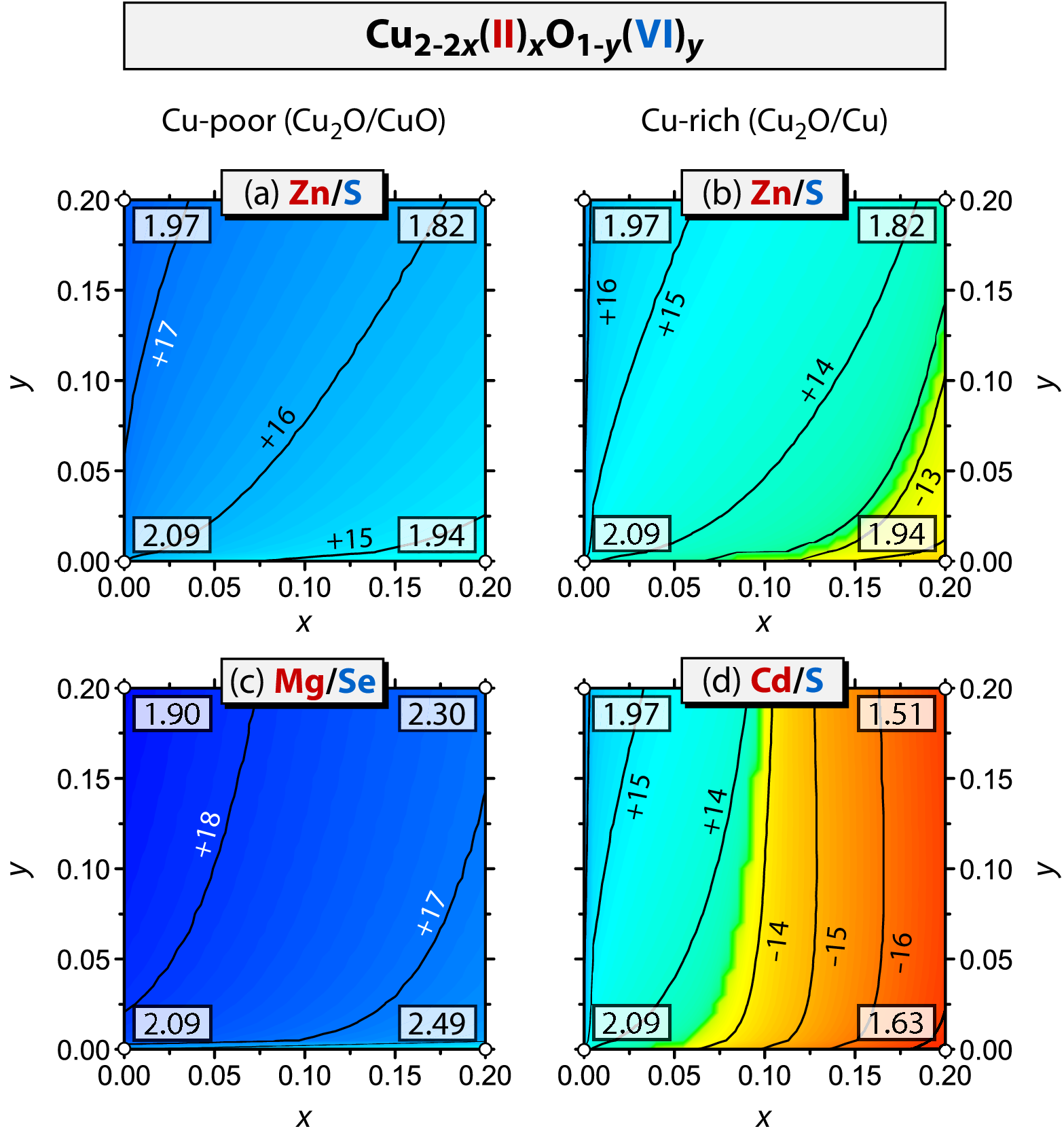}
\caption{\label{fig:Fig1_new}
Thermodynamic modeling (T=400$^{\circ}$ C) of the net doping 
$log(|N_D-N_A|/\mathrm{cm}^{-3})$ in Cu$_{2-2x}$(II)$_x$O$_{1-y}$(VI)$_y$
alloys as  a function of $x$ and $y$ for 4 different II/VI combinations. $N_D$  and $N_A$ are individual concentrations of donors 
and acceptors, respectively. The sign indicates the type of doping (positive for $p$-type, negative for $n$-type). The band gap values
extrapolated according to eq.~\eqref{eq:modeling} are given for the end compositions for $0 \leq (x,y) \leq 0.2$.}
\end{figure}

We approach the  aliovalent alloy problem by extending the conventional dilute impurity model to higher (alloy) concentrations and study 
aliovalent alloying of divalent cations (II = Mg, Zn, Cd) and isovalent 
chalcogenide anions (VI = S, Se) into a Cu$_2$O matrix.
Specifically, we first calculate the formation energies of substitutional dopants and intrinsic defects in the dilute limit. 
Second, we determine the structures and binding energies of dopant-defect pairs and complexes. Third, knowing the 
energetically favorable defect structure, we determine the compositional dependence of the band-gap and band-edge energies, 
which affect the defect formation energies. Finally, based on this input data, we perform thermodynamic simulations for the 
net doping concentrations as a function of the alloy composition. The results of these simulation are shown in Fig.~\ref{fig:Fig1_new}, 
and we will describe the individual steps in detail below. 

The Cu$_2$O parent compound has received 
considerable interest as one of the few prototypical $p$-type oxides \cite{raebiger_PRB:2007, scanlon_PRL:2009,nolan_CM:2008}, 
and the understanding of the band-structure and defect physics in Cu$_2$O is central to succeed in the quest for the so far 
elusive p-type transparent conductive oxides \cite{kawazoe_Nature:1997, hautier_NC:2013}. 
The p-type nature of Cu$_2$O has further spurred interest in the areas of magnetic  
semiconductor \cite{kale_APL:2003,raebiger_PRL:2007} and in regard of possible applications in photovoltaics 
\cite{mittiga_APL:2006,meyer_PSS:2012} and photoelectrocatalysis \cite{paracchino_NM:2011}. However, as in case of the traditional 
semiconductor alloys, the controlled tailoring of the band-structure and electrical properties will be 
instrumental in realizing novel Cu$_2$O based technologies. In particular, ambipolar dopability would open a range of 
potential applications from oxide electronics to solar energy generation.
%
\section*{Approach and results}
%
In order to theoretically predict both band-structure properties and electrical doping as a function of the 
alloy composition, we start from the conventional defect theory and supercell formalism 
\cite{walle_JAP:2004,lany_PRL:2007,lany_PRB:2008,agoston_JAP:2010} 
and then formulate an approach to extend the dilute impurity model to the higher concentrations present in alloys. 
Within the standard dilute defect model, the formation energy of a defect $D$ in a charge state $q$ is defined as
\begin{equation}\label{eq:dhd}
\begin{gathered}
\Delta H_{D,q} (\Delta E_F,\{\Delta \mu_\alpha\}) = [E_{D,q} - E_H] + \\
+ q\,(E_{VBM} + \Delta E_F) + \sum_\alpha n_\alpha (\mu^0_\alpha + \Delta \mu_\alpha),
\end{gathered}
\end{equation}
and is a function of two types of variables: (i) $\Delta E_F$ , measuring the Fermi energy $E_F$ relative to the valence
band maximum (VBM) of the host system and (ii) a set of chemical potentials $\{\Delta \mu_\alpha\}$
describing chemical reservoirs. $E_{D,q}$ and $E_H$ in eq.~\eqref{eq:dhd} are the total energies of a system with and 
without the defect, respectively. The $\{\Delta \mu_\alpha\}$ are defined relative to 
chemical potentials $\{\mu^0_\alpha\}$ of the pure elements in their reference 
phases and reflect the thermodynamic boundary conditions, ranging between Cu-poor / O-rich (CuO/Cu$_2$O coexistence) to  Cu-rich / O-poor (Cu/Cu$_2$O). 
For an accurate prediction of the defect formation energies, we use a recently introduced and tested approach \cite{peng_PRB:2013} that 
combines supercell calculation using density functional theory (DFT) with band-gap corrections from GW quasi-particle energy calculations.
Further details of our computational approach are given in the Methods section.
\begin{figure}[t!]
\centering
\includegraphics[width=\linewidth]{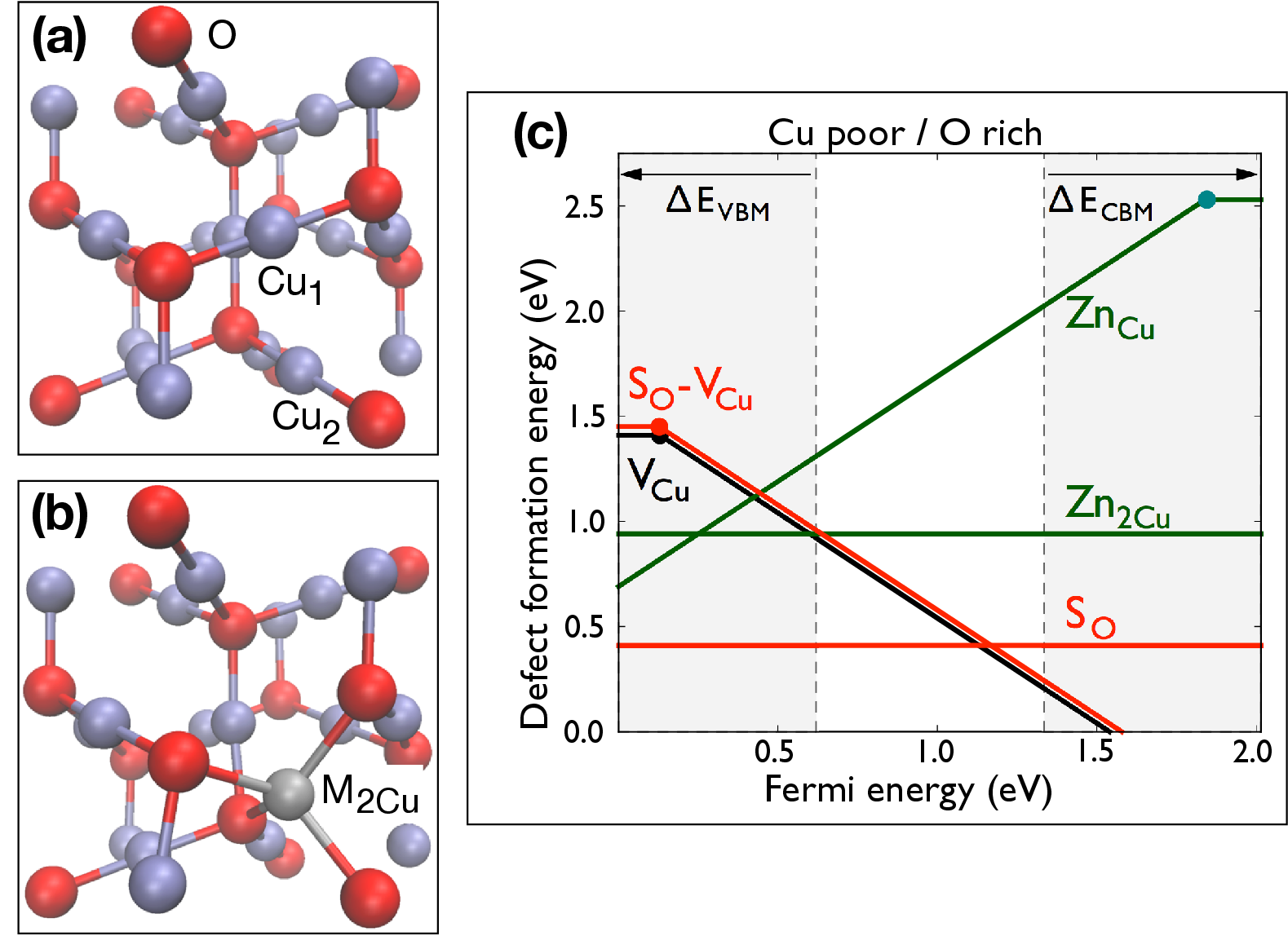}
\caption{\label{fig:Fig2_new}
(a) Cuprite Cu$_2$O structure with O atoms shown in red and Cu in blue; (b) structure of a (II)$_{\mathrm{2Cu}}$ 
defect pair, where one metal impurity (II=Mg,Zn,Cd) shown in grey replaces two copper atoms Cu$_1$ and Cu$_2$;  
(c) defect and defect-pair formation energies of V$_\mathrm{Cu}$, group II cation impurities (II=Zn) and and group VI anion impurities (VI=S) as a function 
of the Fermi energy, assuming phase coexistence of Cu$_2$O with ZnO and Cu$_2$S.  
}
\end{figure}

With an increasing concentration of dopants beyond the dilute limit, two effects become more prominent: First, the interaction between dopants and defects 
can lead to the formation of pairs and larger complexes. Thus, we calculate the different configurations of dopant-defect pairs and their binding energies,
and take into account their association and dissociation within the thermodynamic modeling, using the law of mass action \cite{biswas_PRB:2009}.
Second, since the band-structure changes with the chemical composition, we need to take into account the composition dependence of the 
individual band edge energies ($E_{VBM}$ and $E_{CBM}$), which control the formation energies of ionized (charged) defects and dopants (cf. eq. \eqref{eq:dhd}).   
%
\subsection*{Energetics of point defects and defect pairs}
%
\begin{figure}
\centering
\includegraphics[width=0.8\linewidth]{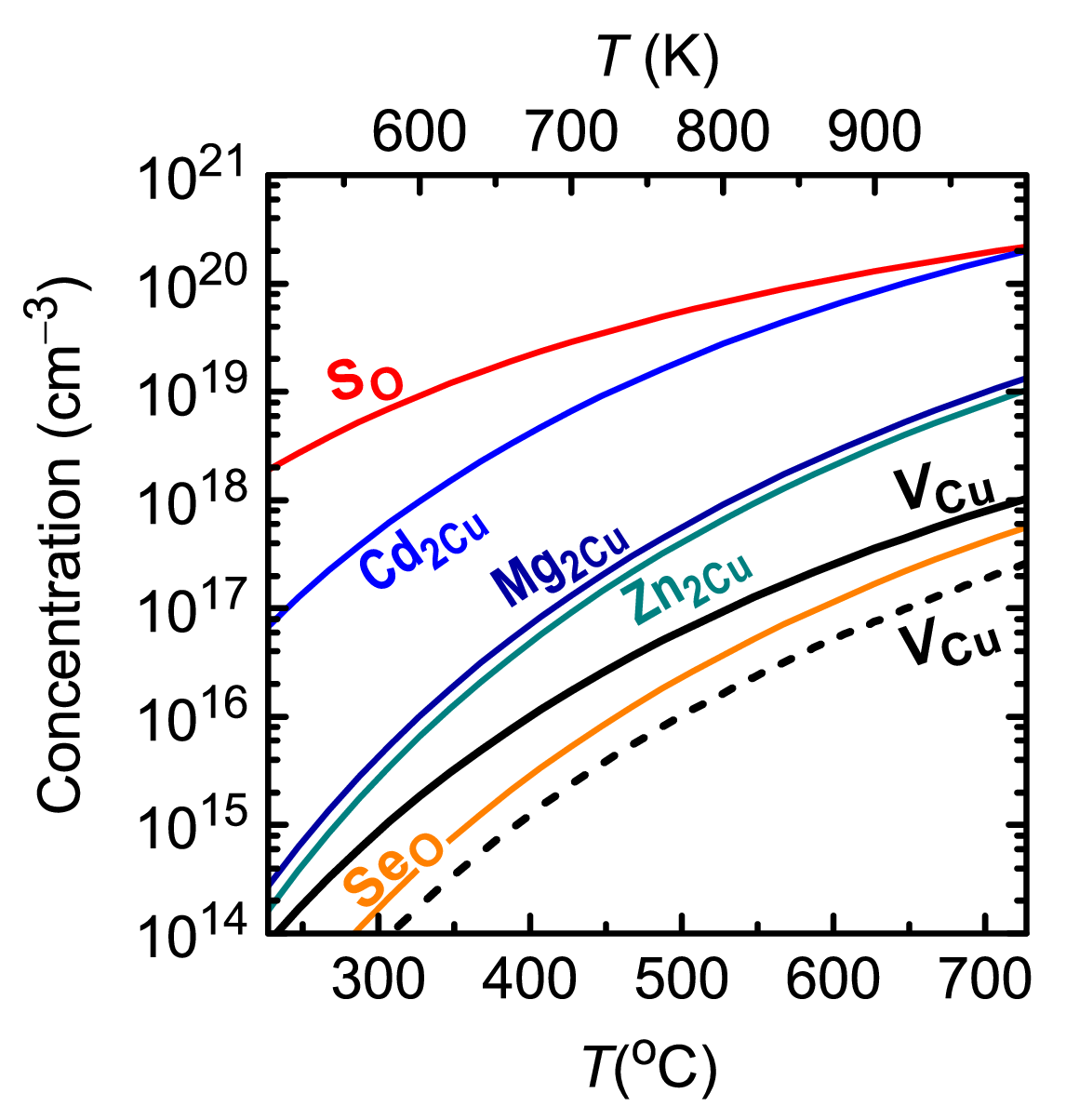}
\caption{\label{fig:Fig3_new}
Thermodynamic modeling of defect and dopant concentrations in Cu$_2$O. Solid lines show 
the solubility limits under Cu-poor (Cu$_2$O/CuO boundary) conditions. The dashed line shows 
the V$_{\mathrm{Cu}}$ concentration under Cu-rich conditions (Cu$_2$O/Cu boundary).}
\end{figure}
In order to develop a complete defect model for the underlying cuprite structure of Cu$_2$O, we consider the intrinsic defects, i.e. the cation and 
anion vacancies and interstitials, the extrinsic cation and anion substitutions, as well as defect pairs and complexes between the 
low-energy species, up to three constituents. 
Fig.~\ref{fig:Fig2_new}(c) shows the defect formation energies of the most relevant point defects and defect pairs, 
as a function of the Fermi energy at the Cu-poor/O-rich conditions and for the particular 
II=Zn and VI=S choice. The chemical potential of Zn is limited by the formation of ZnO.
The other considered cases of II=Mg,Cd and VI=Se present a qualitatively similar picture. 
The full list of calculated formation energies is given in the Appendix~I.
The shaded areas denote the band edge shifts $\Delta$E$_{VBM}$ and $\Delta$E$_{CBM}$ determined from GW calculations. 
Since oxygen vacancies stay in the electrically inactive neutral charge state irrespective of the Fermi level \cite{raebiger_PRB:2007}
and do not show strong binding to other defects, we will not further discuss them.

Zn$_{\mathrm{Cu}}$ is an electrically active donor-type defect (divalent Zn substituting for monovalent Cu) that 
assumes a positively charged state for most Fermi energies, and which has a shallow donor level about 0.18 eV 
below the conduction band minimum. The 
positively charged Zn$^{+}_{\mathrm{Cu}}$ attracts negatively charged V$^{-}_\mathrm{Cu}$ resulting in the formation of the 
electrically neutral Zn$_{2\mathrm{Cu}}$ defect complex in which 
Zn substitutes for two Cu atoms and occupies an interstitial site that is four-fold coordinated by 
oxygen as shown in Fig.~\ref{fig:Fig2_new}(b). 
This configuration is akin to the Cu vacancy in the "split vacancy" configuration (one interstitial Cu replaces two lattice Cu atoms), 
which is a metastable configuration about 0.3 eV higher in energy than the vacancy at the Cu lattice site \cite{wright_JAP:2002, raebiger_PRB:2007}. 
Here, however, the Zn$_{2\mathrm{Cu}}$ configuration is the ground state, which accommodates the preferential tetrahedral 
coordination of Zn inside the cuprite lattice, and lies about 1.29 eV lower in energy than the (Zn$^{+}_\mathrm{Cu}$-V$^{-}_\mathrm{Cu}$) 
pair. Analogous defect complexes are formed by the other group II elements Mg and Cd.
Further, also the isovalent S$_\mathrm{O}$ and Se$_\mathrm{O}$ defects bind Cu vacancies, which can be understood as 
resulting from compensation of the tensile (S$_\mathrm{O}$ and Se$_\mathrm{O}$) and compressive strain (V$_\mathrm{Cu}$) 
induced by the defects. The binding energies relative to the isolated dopants and defects are given in Table~\ref{Table1}.
\begin{table}
\centering
\caption{\label{Table1}
Calculated binding energies of defect pairs formed between the isolated (II) and (VI) dopants and Cu vacancies.}
\begin{tabular}{lc}
\hline\hline
   Defect reaction & Binding energy (eV) \\
\hline
Mg$^{+}_{\mathrm{Cu}}$ + V$^{-}_\mathrm{Cu}$ $\rightarrow$ Mg$_{2\mathrm{Cu}}$ & -2.13 \\
Zn$^{+}_{\mathrm{Cu}}$ + V$^{-}_\mathrm{Cu}$ $\rightarrow$ Zn$_{2\mathrm{Cu}}$ & -1.29 \\
Cd$^{+}_{\mathrm{Cu}}$ + V$^{-}_\mathrm{Cu}$ $\rightarrow$ Cd$_{2\mathrm{Cu}}$ & -1.24 \\
S$_{\mathrm{O}}$ + V$^{-}_\mathrm{Cu}$ $\rightarrow$ (S$_{\mathrm{O}}$-V$_\mathrm{Cu}$)$^{-}$ & -0.37 \\
Se$_{\mathrm{O}}$ + V$^{-}_\mathrm{Cu}$ $\rightarrow$ (Se$_{\mathrm{O}}$-V$_\mathrm{Cu}$)$^{-}$ & -0.72\\
\hline
\end{tabular}
\end{table}

%
\subsection*{Composition dependence of the band-edge energies}
%
In addition to defect-pairing, a second effect that needs to be 
taken into account when extending the dilute defect model to larger concentrations is the composition dependence of the 
band-edge energies. Similar as the GW quasiparticle energy shifts (see Fig.~\ref{fig:Fig2_new}(c)), the charged defect formation energies vary with a change
of the VBM and CBM energies with composition. The band gap and band edge shifts in a Cu$_{2-2x}$(II)$_x$O$_{1-y}$(VI)$_y$ 
alloy can then be expressed by a linear expansion ,
%
\begin{equation}\label{eq:modeling}
E_g(x,y) = E_g^0 + \alpha_g^{II}\, x + \alpha_g^{VI}\, y,
\end{equation}
%
and similar expressions with $ \alpha_{VBM}$ and $ \alpha_{CBM}$ for the individual bend edges. 
The $\alpha$ parameters, determined from GW calculations in supercells containing (II)$_{2Cu}$ and (IV)$_O$ substitutions 
are given in Table~\ref{Table2}. We note that none of the (II) and (VI) dopants introduce resonant 
states close to the band edge energies, which could cause large bowing effects, like, e.g. in N doped GaAs \cite{perkins_PRL:1999}. 
Thus, the linear expansion, eq.~\eqref{eq:modeling}, can be expected to be a good approximation within the low to 
moderate composition range up to $x,y \leq 0.2$ considered here. Having the $\alpha$ parameters allows  
predicting directly the defect formation energies of Cu$_{2-2x}$(II)$_x$O$_{1-y}$(VI)$_y$ alloys using the defect formation 
energies from the pure Cu$_2$O computed from eq.~\eqref{eq:dhd} and the composition dependent band edges from eq.~\eqref{eq:modeling}. 
We tested the applicability and accuracy of this approach by performing direct defect calculations on actual alloy 
compositions as shown in the Appendix~III.
\begin{table}[h!]
\centering
\caption{\label{Table2}
The parameters $\alpha$ (eV) describing the composition dependence of the band-edge and -gap energies, according to eq.~\eqref{eq:modeling}.}
\begin{tabular}{lccc}
\hline\hline
                & $\alpha_{VBM}$ & $\alpha_{CBM}$ & $\alpha_{g}$ \\
\hline
II = Mg   &      -1.88                &        +0.12                      &         +2.00                      \\
II = Zn    &      -0.59                &        -1.32                      &         -0.73                      \\
II = Cd   &      -0.97                &        -3.26                      &         -2.30                      \\
VI = S    &      +0.33               &        -0.30                      &         -0.62                      \\
VI = Se  &      +0.06               &        -0.88                      &         -0.95                      \\ 
\hline
\end{tabular}
\end{table}

\subsection*{Thermodynamic modeling of defect concentrations and doping}
%
Using the calculated formation energies of point defects and defect pairs, as well as their composition dependence, 
we performed thermodynamic simulations to determine the concentrations of the substituted elements and the V$_\mathrm{Cu}$ 
defects. Under equilibrium conditions, the site concentration of a defect $D$, i.e., the concentration of defects divided by 
that of the lattice sites on which the defect resides, is given approximately by $[D] = exp(-\Delta H_D/k_{B}T)$. Due to the Fermi 
level dependence of $\Delta H_D$ in case of electrically active defects or dopants (cf. eq.~\eqref{eq:dhd}), $\Delta E_F$ needs to 
be solved together with the defect concentrations, which is achieved by a numerical self-consistent solution under the constraint of overall 
charge neutrality between charged defects and carriers (free electrons and holes) \cite{lany_APL:2005,lany_PRL:2007}. In the present work, 
we further take into account the association and dissociation of defect pairs within the self-consistent solution, as described 
in Ref.~\cite{biswas_PRB:2009}. 
The thermodynamics of the association and dissociation of defect pairs is described by the law of mass action, e.g.
%
\begin{equation}\label{eq:mass_act}
\begin{gathered}
\left[ \mathrm{Zn}_{2\mathrm{Cu}} \right] =  [\mathrm{Zn}_\mathrm{Cu}] [\mathrm{V}_\mathrm{Cu}] exp(-E_b(\mathrm{Zn}_{2\mathrm{Cu}})/k_BT)),\\
[(\mathrm{S}_\mathrm{O}-\mathrm{V}_\mathrm{Cu})] = [\mathrm{S}_\mathrm{O}] [\mathrm{V}_\mathrm{Cu}] exp(-E_b(\mathrm{S}_\mathrm{O}-\mathrm{V}_\mathrm{Cu})/k_BT);
\end{gathered}
\end{equation}
%
where the brackets denote the site concentrations of the respective species and include the multiplicity of the equivalent
configurations of the defect pairs \cite{biswas_PRB:2009}.
The electrical properties are characterized by the common "net doping concentration", i.e., the 
difference $N_D - N_A$ between the total donor and acceptor concentrations. In the following, we apply this model to three 
situations, (i) the intrinsic doping of pure Cu$_2$O due to V$_\mathrm{Cu}$ formation, (ii) the equilibrium solubility limits of the group II 
and VI dopants, and (iii) the composition dependence of electrical properties in alloys at non-equilibrium compositions.

As shown in Fig.~\ref{fig:Fig3_new}, we find that in pure Cu$_2$O, the V$_\mathrm{Cu}$ concentration varies between the mid-10$^{15}$ to mid 10$^{16}$ cm$^{-3}$ 
range between the Cu-rich (Cu$_2$O/Cu) and Cu-poor (Cu$_2$O/CuO) conditions at T = 450$^\circ$C, agreeing well with the hole carrier density 
of about 10$^{15}$ cm$^{-3}$ measured in Cu$_2$O sheets quenched from this temperature \cite{mittiga_APL:2006,mittiga_TSF:2009}. 
In order to determine the solubility limits of the group II and VI dopants, we take into account the constraints to their chemical potentials arising 
from phase separation and the ensuing precipitation of the competing phases, i.e., MgO, ZnO, CdO, Cu$_2$S, Cu$_2$Se. As seen in Fig.\ref{fig:Fig3_new}, 
the resulting equilibrium solubilities can exceed 10$^{20}$ cm$^{-3}$, but remain in the range of dilute doping below the percent range. 
Since the dominant defect configurations, i.e., the (VI)$_\mathrm{O}$ substitution and the (II)$_\mathrm{2Cu}$ pair, are charge neutral (cf. Fig.~\ref{fig:Fig2_new}c), 
and since the effect on the band energies is minute at such low concentrations, the electrical properties do not significantly change compared to pure Cu$_2$O.  

The solubility limits of dopants can often be overcome by non-equilibrium techniques, such as low-temperature thin-film growth \cite{Desnica_98,zak_PRB:2012}. 
In fact, the solubility limits are often attained only after prolonged annealing procedures at high temperatures \cite{Mason_12}. Similarly, in alloys 
where the positive mixing enthalpy creates a "miscibility gap" in equilibrium, such compositions can nevertheless be realized under synthesis 
conditions where the long-range diffusion necessary for phase separation is kinetically limited \cite{singh_APL:1997}. In case of heterostructural 
alloys, the lattice mismatch provides a further barrier for nucleation of secondary phases. Thus, we are now addressing the question which 
range of band-gaps and electrical doping can be achieved in Cu$_{2-2x}$(II)$_x$O$_{1-y}$(VI)$_y$ if the alloy composition is treated as a parameter 
that can exceed the thermodynamic solubility limit. To this end, we performed the thermodynamic modeling for a partial equilibrium \cite{lany_MSSE:2009}, 
in which the constraints due to phase separation and precipitation are omitted. In practice, the dopant chemical potential is adjusted during 
the thermodynamic simulation until the respective alloy composition is attained. This situation corresponds to a supersaturation of dopants, 
i.e., the dopant chemical potential is higher and the respective defect formation energy is lower than in the unconstrained equilibrium 
where precipitation of secondary phases (e.g., ZnO, Cu$_2$S, etc.) limits the solubility. The balance between electrically active dopants 
and the compensating intrinsic defects determines the electrical properties. Such partial equilibrium simulations have recently explained 
successfully the temperature dependence of the conductivity in Ga doped ZnO \cite{zakutayev_APL:2013}.

The contour plot in Fig.~\ref{fig:Fig1_new} shows the net-doping $log(|N_D-N_A|/\mathrm{cm^{-1}})$
in Cu$_{2-2x}$(II)$_x$O$_{1-y}$(VI)$_y$ alloys as a function of $x$ and $y$ ($N_D$ and $N_A$
stand for the concentration of donors and acceptors, respectively). The numbers in the corners give the predicted band gaps for the respective 
end point compositions according to eq.~\eqref{eq:modeling} and the data in Table~\ref{Table2}. 
For the Zn/S combination, we show the net-doping for both Cu poor (equilibrium between Cu$_2$O and CuO) and Cu rich (equilibrium
between Cu$_2$O and Cu) conditions, thereby illustrating the dependence on the growth conditions. 
The combinations Mg/Se and Cd/S are shown for the Cu poor and Cu rich conditions, respectively, thereby 
emphasizing the most pronounced $p$-type (due to Se) and $n$-type (due to Cd) doping scenarios.
Within the composition range $0 \leq x, y \leq 0.2$, we obtain band gaps between 1.44 eV (for the Cd/Se combination at$x,y = 0.2$, 
not shown in Fig.~\ref{fig:Fig1_new}) and 2.49 eV (for Mg alloying at $x = 0.2$). 
The complete data set including all II/VI combinations is given in the Appendix~II.
\begin{figure}
\centering
\includegraphics[width=\linewidth]{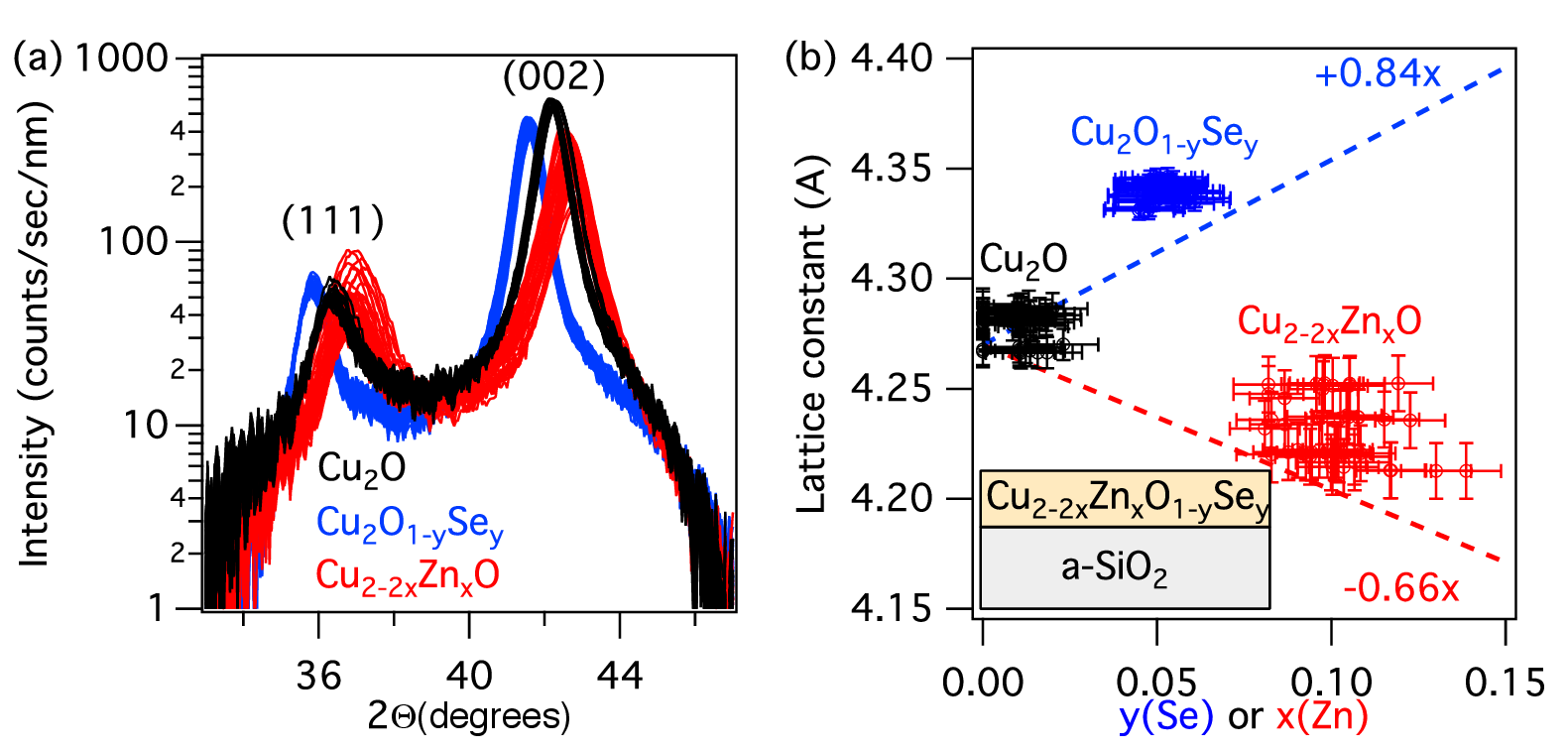}
\caption{\label{fig:Fig4}
(a) X-ray diffraction patterns of Cu$_2$O (black), Cu$_{2-2x}$Zn$_x$O (red) and Cu$_2$O$_{1-y}$Se$_y$ (blue) thin films on a-SiO$_2$, 44 patterns each.
(b) Lattice constant of Cu$_2$O (black), Cu$_{2-2x}$Zn$_x$O (red) and Cu$_2$O$_{1-y}$Se$_y$ (blue) alloys from experiment (symbols) and computations (lines)}
\end{figure}
%
%
\subsection*{Experimental synthesis}
%
As a first step towards the experimental realization of these novel functional Cu$_2$O alloys, we used an combinatorial 
synthesis and characterization approach \cite{zak_RSI:2013} to grow thin films of Cu$_2$O, Cu$_{2-2x}$Zn$_x$O, and Cu$_2$O$_{1-y}$Se$_y$ \cite{zak_MRSC:2011},and to characterize these films in spatially resolved way \cite{zak_PRB:2012} using X-ray diffraction. 
Figure~\ref{fig:Fig4}(a) shows the X-ray diffraction data for pure Cu$_2$O,
for varying Zn substitution centered around $x=0.10$, and for varying Se substitution centered around $y=0.06$.
No impurity phases of ZnO or Cu$_2$Se are observed. The measured 
composition dependence of the lattice constant is compared with the prediction of the defect model in Fig.~\ref{fig:Fig4}(b). The trend of a reduction 
of the lattice constant due to Zn alloying and an increase due to Se alloying is 
observed in both the experimental and theoretical data sets, suggesting that 
the alloyed elements are incorporated in the cuprite 
lattice as described by the computational defect model, instead of forming secondary phases.
Note that the presence of residual strain in the films leads to a slightly larger lattice constant of Cu$_2$O than in single crystals, where $a$ = 4.27 {\AA}, 
and that the present type of DFT calculations cause a typical, but not exactly systematic overestimation of the lattice constant by about 1 \%.
%
\section*{Discussion}
%
The extension of the traditional dilute defect model towards larger alloy concentrations enables the prediction of doping
in aliovalent alloys, thereby opening a path to design the electrical properties in complex semiconductor materials.
Notably, the electrical behavior of the Cu$_{2-2x}$(II)$_x$O$_{1-y}$(VI)$_y$ alloys differs markedly from the traditional doping mechanism 
\cite{woodyard_USP:1950}, where aliovalent impurity atoms introduce a number of charge carriers that is comparable to the number dopant 
atoms (although a certain reduction from unity doping efficiency is often caused by self-compensation \cite{mandel_PR:1964}). 

Our model predicts an interesting and counterintuitive doping behavior, in that the aliovalent group-II dopants have a negligible effect 
on the electrical properties at typical sub-percent doping levels (see Fig.~\ref{fig:Fig1_new}), but modify the band gap at higher alloy-like 
concentrations. On the other hand, the isovalent group-VI substitution has a rather modest effect on the band gap, but increases 
the hole-carrier concentration significantly. This rather ironic behavior is explained by the important role of dopant-defect interactions 
in this system: The divalent dopants incorporate dominantly in the form of a valence-conserving defect complex, e.g., Zn$_\mathrm{2Cu}$, 
where one Zn$^\mathrm{+II}$ replaces two Cu$^\mathrm{+I}$ ions. This charge-neutral complex is electrically inactive, but it modifies 
the band structure. As seen in see Table~\ref{Table2}, Mg alloying lowers the VBM energy, which can be explained by the fact that Mg lacks an 
occupied $d$-shell, and therefore reduces the density of states at the Cu-$d$ like top of the valence band. Zn and Cd introduce delocalized, 
unoccupied $s$-like states, which cause a lowering of the CBM energy. The isovalent dopants have a smaller effect on the band 
structure but affect the electrical properties. The binding energy between substitutional S$_\mathrm{O}$ or Se$_\mathrm{O}$ dopants and 
Cu vacancy V$_\mathrm{Cu}$ effectively reduces the formation enthalpy of these hole-producing defects when forming close 
to the isovalent dopant, thereby increasing the $p$-type doping with increasing S/Se alloying.

We considered the range $0  \leq x,y  \leq 0.2$ as a composition range within which the realization of Cu$_{2-2x}$(II)$_x$O$_{1-y}$(VI)$_y$ 
alloys could be achievable by non-equilibrium growth methods. Within this composition window, we obtain predicted band gap energies between 
1.4 ($x_\mathrm{Cd} = y_\mathrm{Se} = 0.2$) and 2.5 eV ($x_\mathrm{Mg} = 0.2$) from the values given in Table~\ref{Table2}, compared to the 
2.1 eV room temperature gap of Cu$_2$O \cite{malerba_SEMSC:2011}. 
Whereas the binary oxide Cu$_2$O is always $p$-type conducting within a narrow window 
$p =10^{14} - 10^{16} \mathrm{cm}^{-3}$ \cite{meyer_PSS:2012,mittiga_TSF:2009,papadimitriou_SSE:1993}
we find the alloying approach allows a much better control of the electrical properties. 
Due to the pronounced dopant-defect interaction, alloying of S and Se increases the $p$-type doping, up to the $10^{18} \mathrm{cm}^{-3}$ range 
for $y_\mathrm{Se} > 0.02$ (Fig.~\ref{fig:Fig1_new}). For the case of cation doping, the formation of dopant-defect complexes like Zn$_\mathrm{2Cu}$ 
prevents effective $n$-type doping. However, at very high concentrations of Zn or Cd beyond the dilute doping regime, i.e. in the aliovalent 
alloying regime, we observe type conversion from $p$- to $n$-type with a maximal electron doping level of $N_D-N_A = 2\times 10^{17} \mathrm{cm}^{-3}$ 
at $x_\mathrm{Cd} = 0.2$ (see Fig.~\ref{fig:Fig1_new}).

The physical origin of the type conversion from $p$- to $n$-type is a combination of two causes. 
First, the increased non-equilibrium chemical potential of the group II element effectively lowers the formation energy of the (II)$_\mathrm{Cu}$ 
donors, which otherwise is rather high (see Fig.~\ref{fig:Fig2_new}(c)). Note that most II elements are still incorporated as charge neutral (II)$_\mathrm{2Cu}$ 
defect pairs, and only a fraction forms as a substitutional donor, e.g., Cd$_\mathrm{Cu}$, as determined by the law of mass action (see eq.~\eqref{eq:mass_act}). 
In fact, only a fraction of about $10^{-6}$ of alloyed Cd atoms become electrically active as $n$-type dopants. The second effect is the lowering 
of the CBM energy with the $x$ composition for Zn and Cd, which supports $n$-type doping by bringing the CBM closer to the equilibrium 
Fermi level during the thermodynamic simulation. This effect is most pronounced for Cd alloying which affords the largest reduction of the 
CBM energy, as seen in the $\alpha_{CBM}$ parameter in Table~\ref{Table2}, and which is the only group II dopant that can be expected to produce 
robust $n$-type doping with appreciable carrier densities (Fig.~\ref{fig:Fig1_new}). 

The synthesis and characterization 
of Zn and Se substituted Cu$_2$O alloys shows no indication of impurity phases (ZnO or Cu$_2$Se), 
and the composition dependence of the lattice parameter is consistent with the computational defect model. 
Thus, the initial experimental data supports the viability of the proposed alloy system.
%
\section*{Conclusions}
%
In conclusion, the band-structure and electrical properties of complex Cu$_{2-2x}$(II)$_x$O$_{1-y}$(VI)$_y$ alloys 
were modeled by extending the dilute defect model to finite alloy compositions, taking into account pair and complex formation 
between of dopants and defects, as well as the composition dependence of the band edge energies. In contrast to conventional 
semiconductor systems, where the manipulation of band-structure properties via isovalent alloying is separated from the control 
of electrical properties via dilute aliovalent doping, the two mechanisms become intertwined due to the dopant-defect interactions. 
Considering the alloying of aliovalent (Mg, Zn, Cd) cations and isovalent anions (S, Se) into Cu$_2$O, we predicted
that the band-gap energies and the doping levels are tunable over a wide range (gap from 1.4 to 2.5 eV, carriers from $p=10^{18}$ cm$^{-1}$
to $n=2\times 10^{17}$ cm$^{-1}$), including 
the type conversion from p- to n-type. The initial thin film synthesis and characterization of these novel oxide semiconductor materials 
has shown a single phase formation beyond the thermodynamic solubility limit, thereby supporting the underlying defect model. 
The Cu$_{2-2x}$(II)$_x$O$_{1-y}$(VI)$_y$ alloys could find application, e.g., as alternative earth abundant photovoltaic materials. 
%
%
\section*{Acknowledgements}
%
This work was supported by the US Department of energy, Office of Energy Efficiency and Renewable 
Energy, under Contract No. DE-AC36-08GO28308 to the National Renewable Energy Laboratory (NREL), as part of a
Next Generation Photovoltaics II project within the SunShot initiative. The use of high performance computing 
resources of the National Energy Research Scientific Computing Center and of NREL's Computational Science Center are 
gratefully acknowledged. V.S. also acknowledges the administrative support of REMRSEC at the Colorado School of Mines.
%
\section*{Appendix I: Methods}
%
Theory and modeling: All DFT calculations were performed with the projector augmented-wave (PAW) method \cite{blochl_PRB:1994} 
as implemented in the VASP code \cite{joubert_PRB:1999}, employing the PBE exchange-correlation functional \cite{perdew_PRL:1996}, 
and the DFT+U formulation of Ref.~\cite{dudarev_PRB:1998}, with U = 5 eV for Cu-$d$ orbitals. Defects and defect pairs 
were modeled in large supercells of Cu$_2$O with 162 atoms, applying the standard finite-size corrections for image charge interactions 
and potential alignment \cite{lany_MSSE:2009}. The band gap problem was resolved by combining the DFT supercell energies with the results
from GW quasi-particle energy calculations as described in Ref.~\cite{peng_PRB:2013}, using the band edge shifts 
$\Delta$E$_{VBM} = -0.62$ eV and $\Delta$E$_{CBM} = +0.68$ eV, as determined in a recent GW study of transition
metal oxides \cite{lany_3dGW:2013}. In oder to accurately describe the chemical potentials $\{\Delta \mu_\alpha\}$ 
entering in eq.~\eqref{eq:dhd}, we have used the fitted elemental refrence energies (FERE) of Ref.~\cite{vladan_FERE}. 
For the host atoms Cu and O, the chemical potentials are limited by the phase coexistence of Cu$_2$O with CuO 
(Cu-poor/O-rich condition; $\Delta \mu_{\mathrm{Cu}}=-0.24$ eV and $\Delta \mu_{\mathrm{O}}=-1.38$ eV) and with metallic Cu 
(Cu-rich/O-poor condition; $\Delta \mu_{\mathrm{Cu}}= 0.00$ eV and $\Delta \mu_{\mathrm{O}}=-1.86$ eV).
For the thermodynamic solubility limits, the chemical potentials $\{\Delta \mu_\alpha\}$ of the extrinsic impurities are 
determined by the condition of phase coexistence with the related secondary phases, e.g. MgO, ZnO, CdO, Cu$_2$S, Cu$_2$Se.
All defect formation energies for the charge neutral defects and defect pairs, given for $\Delta\mu_{\alpha} = 0$, are provided in Table~\ref{Table_app_2}

To determine the effect of (II)$_{2\mathrm{Cu}}$ and (VI)$_\mathrm{O}$ substitution on band-edge energies, we performed 
direct GW calculations in 48 atom supercells containing one of these defects. These GW calculations were performed analogously to those in 
Ref.~\cite{lany_3dGW:2013} using the implementation of the GW method in the PAW framework Ref.~\cite{shishkin_gwPRB:2006}. 

We note that GW calculations are generally difficult to converge \cite{samsonidze_PRL:2011,friedrich_PRB:2011}. 
For the PAW implementation of the GW method, a recent work  \cite{klimes_arx:2014} has identified limitations due to basis set incompleteness, 
which are particularly pronounced for d-orbitals. We think it is likely that these issues lie behind the previously observed need to apply an external 
d-state potential in GW for transition metal compounds \cite{lany_3dGW:2013}. Including these potentials (here, V$_d$ = -2.4, -1.5, and -0.5 eV 
for Cu, Zn, and Cd, respectively) mitigates these issues, and should lead to fairly reliable valence band shifts, as indicated by the good agreement 
of the calculated ionization potentials with experimental data \cite{stevanovic_PCCP:2013}.       

We further performed a test of the underlying assumption of our model that the charged defect formation enthalpies 
vary linearly with the change of the VBM energy (cf. eqs. \eqref{eq:dhd} and \eqref{eq:modeling}). To this end, we performed defect calculations in
explicit alloy supercells using the same 162 atom Cu$_2$O cells as above, sampling over different alloy compositions (4 and 8 cation or anion substitutions),
alloy configurations and different defect sites in each representation. Since GW calculations for such large supercells are not feasible, 
this test was performed on the DFT+U level. Further details are provided in Supplementary materials. 

In order to calculate the doping and defect concentrations, we use a thermodynamic model  \cite{lany_PRL:2007,biswas_PRB:2009,lany_APL:2005}, 
where a self-consistency condition is solved numerically for the formation energy $\Delta H_{D,q}$, the defect concentration, and the Fermi level $\Delta E_F$ 
under the constraint of overall charge neutrality. The case of the partial equilibrium is solved by adjusting the defect formation energy for 
atomic substitution during the simulation until the respective alloy concentration is obtained. The temperature dependence of 
the Cu$_2$O band gap, as determined in Ref.~\cite{iwamitsu_PSSC:2012} was taken into account in the thermodynamic simulation. 
The calculated carrier densities in pure Cu$_2$O are in agreement with available experimental data \cite{mittiga_APL:2006,mittiga_TSF:2009}, 
but about 2 orders of magnitude lower than those determined before in Ref.~\cite{raebiger_PRB:2007}. This difference results mostly 
from the GW quasi-particle energy shift of the VBM, which increases $\Delta$H(V$_\mathrm{Cu}^-$) by 0.62 eV relative to a 
standard DFT+U calculation. 

The case of the partial equilibrium is solved by adjusting the defect formation energy for atomic substitution during the 
simulation until the respective alloy concentration is obtained.

Thin-film deposition: Thin films of Cu$_2$O, Cu$_{2-2x}$Zn$_x$O, and Cu$_2$O$_{1-y}$Se$_y$ were grown at ambient temperature 
by combinatorial RF co-sputtering with a continuous composition spread \cite{zak_MRSC:2011} in a AJA International vacuum chamber 
with 10$^{-10}$ atm base pressure, and filled with 10$^{-6}$ atm of ultra high purity Ar. We used 50x50 mm Eagle-XG glass substrates and 
50 mm diameter targets of Cu$_2$O, ZnO and Cu$_2$Se. The films were characterized at 44 spatially distinct locations \cite{zak_PRB:2012},
determining the composition and thickness (350-650 nm range) by x-ray fluorescence (XRF), and determining the phase composition and 
lattice constant by X-ray diffraction (XRD).

\begin{table}
\centering
\caption{\label{Table_app_2}
Calculated formation energies $\Delta H_D$ for the charge neutral defects and defect pairs, given for $\Delta\mu_{\alpha} = 0$, 
i.e., all chemical potentials set at the elemental reference phase. For the electrically active defects, the respective donor (D) or 
acceptor (A) ionization energies ($\varepsilon_D/\varepsilon_A$) are also given.}
\begin{ruledtabular}
\begin{tabular}{lccc}
%
                &                & $\Delta H_D$ [eV] & $\varepsilon_D/\varepsilon_A$ [eV]  \\
\hline
V$_\mathrm{Cu}$                                 &  (A) &  +1.65 &  0.13 \\
Mg$_\mathrm{Cu}$                              &  (D) &  -1.32   &  0.17 \\
Zn$_\mathrm{Cu}$                               &  (D) &  +0.59  &  0.18 \\
Cd$_\mathrm{Cu}$                               &  (D) &  +1.17  &  0.18 \\
V$_\mathrm{O}$                                   &        &  +2.42  &  -  \\
S$_\mathrm{O}$                                   &        &  +1.27  &  -  \\
Se$_\mathrm{O}$                                 &        &  +2.14  &  -  \\
\hline
Mg$_\mathrm{2Cu}$                             &        &  -3.52  &  -  \\
Zn$_\mathrm{2Cu}$                              &        &  -0.76  &  -  \\
Cd$_\mathrm{2Cu}$                              &        &  -0.13  &  -  \\
(S$_\mathrm{O}$-V$_\mathrm{Cu}$)    &  (A) &  +2.55 & 0.13 \\
(Se$_\mathrm{O}$-V$_\mathrm{Cu}$)  &  (A) &  +3.07 & 0.13 \\
\end{tabular}
\end{ruledtabular}
\end{table}
%

\newpage

\section*{Appendix II: Complete data of thermodynamic modeling}

\begin{figure}[h!]
\centering
\includegraphics[width=\linewidth]{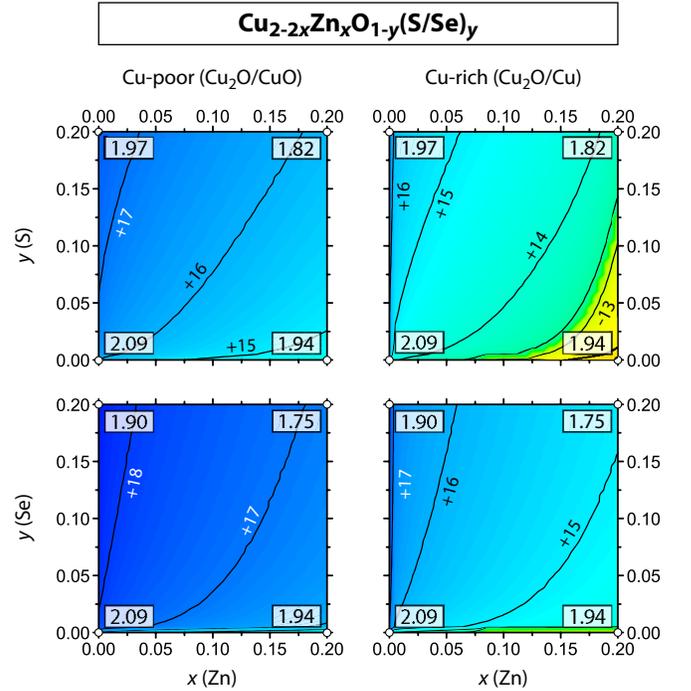}
\caption{\label{fig:Fig7}
Same as Fig.~\ref{fig:Fig1_new}, but for (II) = Zn.}
\end{figure}
\begin{figure}[h!]
\centering
\includegraphics[width=\linewidth]{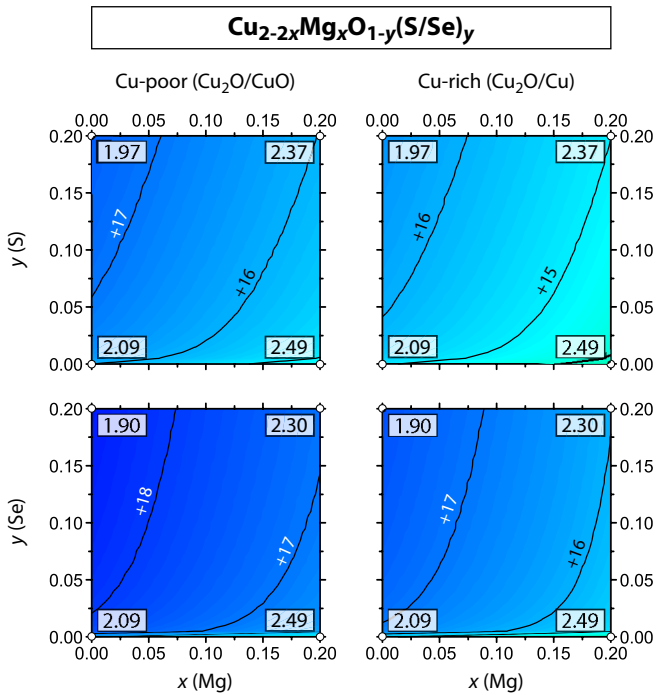}
\caption{\label{fig:Fig7}
Same as Fig.~\ref{fig:Fig1_new}, but for (II) = Mg.}
\end{figure}
\begin{figure}[h!]
\centering
\includegraphics[width=\linewidth]{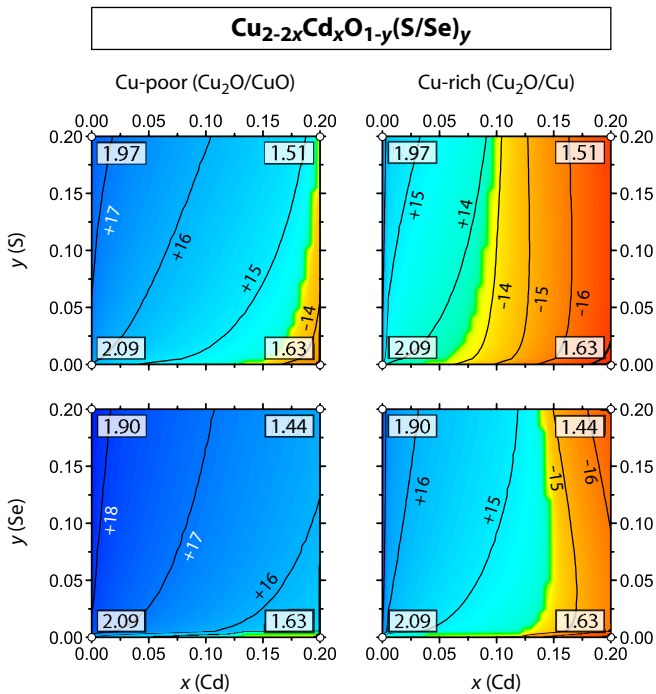}
\caption{\label{fig:Fig7}
Same as Fig.~\ref{fig:Fig1_new}, but for (II) = Cd.}
\end{figure}
%
%
\newpage
%
\section*{Appendix III: Test of alloy model}
%
The present work utilizes a model to determine Ð to linear order Ð the composition-dependence of the band gap and the 
individual band edge energies. We here test this linear extrapolation by comparison with calculations of supercells that 
explicitly incorporate different alloy compositions. Specifically, we are using 162 atom supercells containing 4 and 8 cation 
or anion substitutions, e.g., Zn$_\mathrm{2Cu}$ or S$_\mathrm{O}$ with $x$ or $y$ = 0.074 and 0.148. In all cases, the 
results of two different random alloy representations were averaged. Since GW calculations for such large supercells are 
not feasible, we conduct the test of the model on the GGA+U level. The actual results take also into account 
that the GW quasi-particle energy shifts vary with composition, thereby giving rise to a GW contribution to the composition 
dependence.
\begin{table*}
\centering
\caption{\label{Table_app_1}
The parameters $\alpha$ (eV) describing the composition dependence of the band edge energies, according to eq. \eqref{eq:modeling}. 
Values are given for the DFT contribution (GGA+U), and the contribution from GW quasiparticle energy corrections. The sum of the 
respective contributions yields the values given in Table~\ref{Table2}.}
\begin{tabular}{l|cccc}
\hline\hline
                & $\alpha_{VBM}$(GGA+U) & $\alpha_{CBM}$(GGA+U) &  $\alpha_{VBM}$(GW) & $\alpha_{CBM}$(GW) \\
\hline

Mg$_\mathrm{2Cu}$ &   -1.13             &      +0.40                            &      -0.75                        &     -0.28                        \\
Zn$_\mathrm{2Cu}$ &    -0.94             &      -0.97                             &      +0.36                        &    -0.35                        \\
Cd$_\mathrm{2Cu}$ &   -1.08              &     -2.44                              &      +0.11                       &     -0.82                       \\
S$_\mathrm{O}$       &   -0.76              &     +0.09                             &      +1.09                       &     -0.39                        \\
Se$_\mathrm{O}$     &   -1.07              &     -0.49                              &      +1.13                       &     -0.39                         \\
\hline
\end{tabular}
\end{table*}

\subsection{Composition dependence of the band edge energies}
\begin{figure*}
\centering
\includegraphics[width=0.7\linewidth]{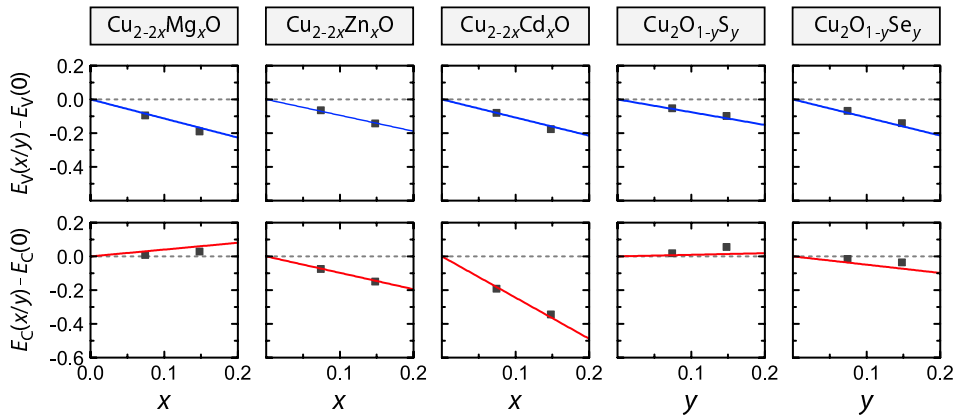}
\caption{\label{fig:Fig5}
Composition dependence of the VBM (top row) and CBM (bottom row) energies. Solid lines represent the extrapolation 
according to eq.\eqref{eq:modeling} with GGA+U parameters from Table~\ref{Table_app_1}, and the data points are results 
from explicit alloy supercells.}
\end{figure*}

In semiconductor alloys, the composition dependence of the band gap and of the band edge energies is usually 
described up to quadratic order via a bowing parameter. However, in the composition window $0 \leq x \leq 0.2$, 
considered in the present work, this dependence is approximately linear. We determined the composition dependence 
of $E_\mathrm{VBM}$, $E_\mathrm{CBM}$, and $E_g$ (see Table~\ref{Table2}) by calculating a single substitution in 
a 48-atom supercell, both in GGA+U and in GW. The individual contributions are given in Table~\ref{Table_app_1}. 
Potential alignment effects have been taken into account to determine the energy lineup between pure and substituted 
Cu$_2$O supercells, using all atoms except the substituted ones as references. The change of the crystal volume with 
composition was taken into account in both the linear extrapolation model and the explicit alloy supercell calculations. 
Figure~\ref{fig:Fig5} shows that both approaches agree very well on the predicted composition dependence of the individual 
band edge energies (and, hence, on the predicted band gap). We note that a deviation from the linear proportionality 
would be expected in case of substitutions that cause defect states inside the band gap, which would create a discontinuity 
of the band edge energies at $x/y > 0$. This behavior is not observed for the alloy substitutions considered in the present work.

\subsection{Composition dependence of the charged-defect formation energies}
\begin{figure*}
\centering
\includegraphics[width=0.7\linewidth]{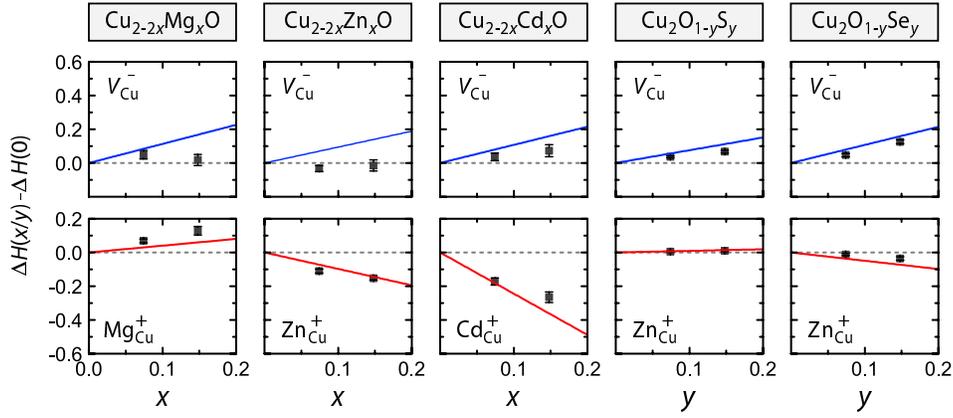}
\caption{\label{fig:Fig6}
Composition dependence of the defect formation energy $\Delta H_{D,q} (E_F)$ of the negatively charged V$_\mathrm{Cu}$ 
defect (top row) and of the positively charged substitutional cation-site donor (bottom row). The formation energy difference 
is evaluated for $E_F = E_\mathrm{VBM}$ in case of the acceptor-like V$_\mathrm{Cu}$ defect and for $E_F = E_CBM$ 
in case of the donor-like cation-site dopant. Solid lines correspond to the extrapolation model using the GGA+U parameters 
from Table~\ref{Table_app_1}, and data points are from defect calculations in explicit alloy supercells.}
\end{figure*}

In order to extend the dilute defect model to low and moderate alloy compositions, we consider two effects that affect 
the defect concentrations. First, the effect of defect pair association is taken into account by calculating the binding energy 
and using the law of mass action. Second, the linear extrapolation of the band edge energies leads to a composition 
dependence of the charged defect ($q \neq 0$) formation energy via eq.~\eqref{eq:dhd}. We test this model by comparison 
with defect formation energy calculations in the explicit alloy supercells, averaging over five different defect sites in each of the two 
alloy representations. Here, we exclude the Cu sites next to the anion (S$\mathrm{O}$ or Se$_\mathrm{O}$) dopants, because 
the treatment of the pair association already accounts for the lowering of the V$_\mathrm{Cu}$ formation energy at these sites. 
Showing the comparison between the extrapolation model and the explicit ally supercell defect calculations, we see in 
Fig.~\ref{fig:Fig6} that for all cases except V$_\mathrm{Cu}$ in the cation-substituted case the model captures well the trends 
in the composition dependence of $\Delta H_{D}$ and gives reasonable quantitative estimates. In the cation substituted cases, 
$\Delta H(\mathrm{V}_\mathrm{Cu})$ calculated in the alloy supercells is lower than the value expected from the extrapolation model. 
This observation can be explained by the fact that that there is a binding energy of, e.g., -0.20 eV between the Zn$_\mathrm{2Cu}$ 
substitution and V$_\mathrm{Cu}$ at the Cu site nearest to the Zn location. 
In principle, one can refine the treatment of the pair and complex association to include larger clusters and more configurations 
with their individual binding energies, e.g., by including a (Zn$_\mathrm{2Cu}$-V$_\mathrm{Cu}$) complex. However, in practice, 
one has to cut off the defect interactions at some point, and we feel that the purpose of the present work is better served by including for clarity 
only the leading mechanisms for defect pair formation that are shown in Table~\ref{Table1}. To conclude, the test using explicit alloy 
supercells has confirmed that the defect pair association and the shift of the band edge energies are the leading effects 
that need to be included to predict defect formation beyond the dilute limit in low and moderate alloy compositions.
%

\begin{thebibliography}{57}%
\makeatletter
\providecommand \@ifxundefined [1]{%
 \@ifx{#1\undefined}
}%
\providecommand \@ifnum [1]{%
 \ifnum #1\expandafter \@firstoftwo
 \else \expandafter \@secondoftwo
 \fi
}%
\providecommand \@ifx [1]{%
 \ifx #1\expandafter \@firstoftwo
 \else \expandafter \@secondoftwo
 \fi
}%
\providecommand \natexlab [1]{#1}%
\providecommand \enquote  [1]{``#1''}%
\providecommand \bibnamefont  [1]{#1}%
\providecommand \bibfnamefont [1]{#1}%
\providecommand \citenamefont [1]{#1}%
\providecommand \href@noop [0]{\@secondoftwo}%
\providecommand \href [0]{\begingroup \@sanitize@url \@href}%
\providecommand \@href[1]{\@@startlink{#1}\@@href}%
\providecommand \@@href[1]{\endgroup#1\@@endlink}%
\providecommand \@sanitize@url [0]{\catcode `\\12\catcode `\$12\catcode
  `\&12\catcode `\#12\catcode `\^12\catcode `\_12\catcode `\%12\relax}%
\providecommand \@@startlink[1]{}%
\providecommand \@@endlink[0]{}%
\providecommand \url  [0]{\begingroup\@sanitize@url \@url }%
\providecommand \@url [1]{\endgroup\@href {#1}{\urlprefix }}%
\providecommand \urlprefix  [0]{URL }%
\providecommand \Eprint [0]{\href }%
\providecommand \doibase [0]{http://dx.doi.org/}%
\providecommand \selectlanguage [0]{\@gobble}%
\providecommand \bibinfo  [0]{\@secondoftwo}%
\providecommand \bibfield  [0]{\@secondoftwo}%
\providecommand \translation [1]{[#1]}%
\providecommand \BibitemOpen [0]{}%
\providecommand \bibitemStop [0]{}%
\providecommand \bibitemNoStop [0]{.\EOS\space}%
\providecommand \EOS [0]{\spacefactor3000\relax}%
\providecommand \BibitemShut  [1]{\csname bibitem#1\endcsname}%
\let\auto@bib@innerbib\@empty
\bibitem [{\citenamefont {Brunner}(2002)}]{brunner_RPP:2002}%
  \BibitemOpen
  \bibfield  {author} {\bibinfo {author} {\bibfnamefont {K.}~\bibnamefont
  {Brunner}},\ }\href@noop {} {\bibfield  {journal} {\bibinfo  {journal} {Rep.
  Prog. Phys.}\ }\textbf {\bibinfo {volume} {65}},\ \bibinfo {pages} {27}
  (\bibinfo {year} {2002})}\BibitemShut {NoStop}%
\bibitem [{\citenamefont {d'Avezac}\ \emph {et~al.}(2012)\citenamefont
  {d'Avezac}, \citenamefont {Luo}, \citenamefont {Chanier},\ and\ \citenamefont
  {Zunger}}]{mayeul_PRL:2012}%
  \BibitemOpen
  \bibfield  {author} {\bibinfo {author} {\bibfnamefont {M.}~\bibnamefont
  {d'Avezac}}, \bibinfo {author} {\bibfnamefont {J.-W.}\ \bibnamefont {Luo}},
  \bibinfo {author} {\bibfnamefont {T.}~\bibnamefont {Chanier}}, \ and\
  \bibinfo {author} {\bibfnamefont {A.}~\bibnamefont {Zunger}},\ }\href@noop {}
  {\bibfield  {journal} {\bibinfo  {journal} {Phys. Rev. Lett.}\ }\textbf
  {\bibinfo {volume} {108}},\ \bibinfo {pages} {027401} (\bibinfo {year}
  {2012})}\BibitemShut {NoStop}%
\bibitem [{\citenamefont {Nakamura}\ \emph {et~al.}(1995)\citenamefont
  {Nakamura}, \citenamefont {Senoh}, \citenamefont {Iwasa},\ and\ \citenamefont
  {Nagahama}}]{nakamura_JJAP:1995}%
  \BibitemOpen
  \bibfield  {author} {\bibinfo {author} {\bibfnamefont {S.}~\bibnamefont
  {Nakamura}}, \bibinfo {author} {\bibfnamefont {M.}~\bibnamefont {Senoh}},
  \bibinfo {author} {\bibfnamefont {N.}~\bibnamefont {Iwasa}}, \ and\ \bibinfo
  {author} {\bibfnamefont {S.}~\bibnamefont {Nagahama}},\ }\href@noop {}
  {\bibfield  {journal} {\bibinfo  {journal} {Jpn. J. Appl. Phys.}\ }\textbf
  {\bibinfo {volume} {34}},\ \bibinfo {pages} {L797} (\bibinfo {year}
  {1995})}\BibitemShut {NoStop}%
\bibitem [{\citenamefont {Knoll}(1999)}]{knoll:1999}%
  \BibitemOpen
  \bibfield  {author} {\bibinfo {author} {\bibfnamefont {G.~F.}\ \bibnamefont
  {Knoll}},\ }\href@noop {} {\emph {\bibinfo {title} {Radiation Detection and
  Measurement}}},\ \bibinfo {edition} {3rd}\ ed.\ (\bibinfo  {publisher}
  {Wiley},\ \bibinfo {address} {New York, USA},\ \bibinfo {year}
  {1999})\BibitemShut {NoStop}%
\bibitem [{\citenamefont {Kinsey}(2010)}]{kinsey:2010}%
  \BibitemOpen
  \bibfield  {author} {\bibinfo {author} {\bibfnamefont {G.}~\bibnamefont
  {Kinsey}},\ }\href@noop {} {\emph {\bibinfo {title} {High-Concentration,
  III--V Multijunction Solar Cells in Solar cells and their applications}}}\
  (\bibinfo  {publisher} {Wiley},\ \bibinfo {address} {New York},\ \bibinfo
  {year} {2010})\BibitemShut {NoStop}%
\bibitem [{\citenamefont {Woodyard}(1950)}]{woodyard_USP:1950}%
  \BibitemOpen
  \bibfield  {author} {\bibinfo {author} {\bibfnamefont {J.}~\bibnamefont
  {Woodyard}},\ }\href@noop {} {\bibfield  {journal} {\bibinfo  {journal} {U.S.
  Patent No. 2,530,110}\ } (\bibinfo {year} {1950})}\BibitemShut {NoStop}%
\bibitem [{\citenamefont {Ginley}\ and\ \citenamefont
  {in}(2010)}]{ginley:2010}%
  \BibitemOpen
  \bibfield  {author} {\bibinfo {author} {\bibfnamefont {D.~S.}\ \bibnamefont
  {Ginley}}\ and\ \bibinfo {author} {\bibfnamefont {J.~D.~P.}\ \bibnamefont
  {in}},\ }\href@noop {} {\emph {\bibinfo {title} {"Handbook of Transparent
  Conductors", D. S. Ginley, H. Hosono, and D. C. Payne}}}\ (\bibinfo
  {publisher} {Springer},\ \bibinfo {address} {New York},\ \bibinfo {year}
  {2010})\BibitemShut {NoStop}%
\bibitem [{\citenamefont {Lindsay}\ and\ \citenamefont
  {O'Reilly}(1999)}]{lindsay_SSC:1999}%
  \BibitemOpen
  \bibfield  {author} {\bibinfo {author} {\bibfnamefont {A.}~\bibnamefont
  {Lindsay}}\ and\ \bibinfo {author} {\bibfnamefont {E.}~\bibnamefont
  {O'Reilly}},\ }\href@noop {} {\bibfield  {journal} {\bibinfo  {journal}
  {Solid State Commun.}\ }\textbf {\bibinfo {volume} {112}},\ \bibinfo {pages}
  {443} (\bibinfo {year} {1999})}\BibitemShut {NoStop}%
\bibitem [{\citenamefont {Kent}\ and\ \citenamefont
  {Zunger}(2001)}]{kent_PRB:2001}%
  \BibitemOpen
  \bibfield  {author} {\bibinfo {author} {\bibfnamefont {P.~R.~C.}\
  \bibnamefont {Kent}}\ and\ \bibinfo {author} {\bibfnamefont {A.}~\bibnamefont
  {Zunger}},\ }\href@noop {} {\bibfield  {journal} {\bibinfo  {journal} {Phys.
  Rev. B}\ }\textbf {\bibinfo {volume} {64}},\ \bibinfo {pages} {115208}
  (\bibinfo {year} {2001})}\BibitemShut {NoStop}%
\bibitem [{\citenamefont {Popescu}\ and\ \citenamefont
  {Zunger}(2010)}]{popescu_PRL:2010}%
  \BibitemOpen
  \bibfield  {author} {\bibinfo {author} {\bibfnamefont {V.}~\bibnamefont
  {Popescu}}\ and\ \bibinfo {author} {\bibfnamefont {A.}~\bibnamefont
  {Zunger}},\ }\href@noop {} {\bibfield  {journal} {\bibinfo  {journal} {Phys.
  Rev. Lett.}\ }\textbf {\bibinfo {volume} {104}},\ \bibinfo {pages} {236403}
  (\bibinfo {year} {2010})}\BibitemShut {NoStop}%
\bibitem [{\citenamefont {Northrup}\ and\ \citenamefont
  {Zhang}(1993)}]{northrup_PRB:1993}%
  \BibitemOpen
  \bibfield  {author} {\bibinfo {author} {\bibfnamefont {J.~E.}\ \bibnamefont
  {Northrup}}\ and\ \bibinfo {author} {\bibfnamefont {S.~B.}\ \bibnamefont
  {Zhang}},\ }\href@noop {} {\bibfield  {journal} {\bibinfo  {journal} {Phys.
  Rev. B}\ }\textbf {\bibinfo {volume} {47}},\ \bibinfo {pages} {6791}
  (\bibinfo {year} {1993})}\BibitemShut {NoStop}%
\bibitem [{\citenamefont {Lany}\ and\ \citenamefont
  {Zunger}(2010)}]{lany_APL:2010}%
  \BibitemOpen
  \bibfield  {author} {\bibinfo {author} {\bibfnamefont {S.}~\bibnamefont
  {Lany}}\ and\ \bibinfo {author} {\bibfnamefont {A.}~\bibnamefont {Zunger}},\
  }\href@noop {} {\bibfield  {journal} {\bibinfo  {journal} {Appl. Phys.
  Lett.}\ }\textbf {\bibinfo {volume} {96}},\ \bibinfo {pages} {142114}
  (\bibinfo {year} {2010})}\BibitemShut {NoStop}%
\bibitem [{\citenamefont {Varley}\ \emph {et~al.}(2010)\citenamefont {Varley},
  \citenamefont {Janotti},\ and\ \citenamefont {de~Walle}}]{varley_PRB:2010}%
  \BibitemOpen
  \bibfield  {author} {\bibinfo {author} {\bibfnamefont {J.~B.}\ \bibnamefont
  {Varley}}, \bibinfo {author} {\bibfnamefont {A.}~\bibnamefont {Janotti}}, \
  and\ \bibinfo {author} {\bibfnamefont {C.~G.~V.}\ \bibnamefont {de~Walle}},\
  }\href@noop {} {\bibfield  {journal} {\bibinfo  {journal} {Phys. Rev. B}\
  }\textbf {\bibinfo {volume} {81}},\ \bibinfo {pages} {245216} (\bibinfo
  {year} {2010})}\BibitemShut {NoStop}%
\bibitem [{\citenamefont {van~de Walle}\ and\ \citenamefont
  {Ellis}(2007)}]{awalle_PRL:2007}%
  \BibitemOpen
  \bibfield  {author} {\bibinfo {author} {\bibfnamefont {A.}~\bibnamefont
  {van~de Walle}}\ and\ \bibinfo {author} {\bibfnamefont {D.~E.}\ \bibnamefont
  {Ellis}},\ }\href@noop {} {\bibfield  {journal} {\bibinfo  {journal} {Phys.
  Rev. Lett.}\ }\textbf {\bibinfo {volume} {98}},\ \bibinfo {pages} {266101}
  (\bibinfo {year} {2007})}\BibitemShut {NoStop}%
\bibitem [{\citenamefont {Raebiger}\ \emph
  {et~al.}(2007{\natexlab{a}})\citenamefont {Raebiger}, \citenamefont {Lany},\
  and\ \citenamefont {Zunger}}]{raebiger_PRB:2007}%
  \BibitemOpen
  \bibfield  {author} {\bibinfo {author} {\bibfnamefont {H.}~\bibnamefont
  {Raebiger}}, \bibinfo {author} {\bibfnamefont {S.}~\bibnamefont {Lany}}, \
  and\ \bibinfo {author} {\bibfnamefont {A.}~\bibnamefont {Zunger}},\
  }\href@noop {} {\bibfield  {journal} {\bibinfo  {journal} {Phys. Rev. B}\
  }\textbf {\bibinfo {volume} {76}},\ \bibinfo {pages} {045209} (\bibinfo
  {year} {2007}{\natexlab{a}})}\BibitemShut {NoStop}%
\bibitem [{\citenamefont {Scanlon}\ \emph {et~al.}(2009)\citenamefont
  {Scanlon}, \citenamefont {Morgan}, \citenamefont {Watson},\ and\
  \citenamefont {Walsh}}]{scanlon_PRL:2009}%
  \BibitemOpen
  \bibfield  {author} {\bibinfo {author} {\bibfnamefont {D.~O.}\ \bibnamefont
  {Scanlon}}, \bibinfo {author} {\bibfnamefont {B.~J.}\ \bibnamefont {Morgan}},
  \bibinfo {author} {\bibfnamefont {G.~W.}\ \bibnamefont {Watson}}, \ and\
  \bibinfo {author} {\bibfnamefont {A.}~\bibnamefont {Walsh}},\ }\href@noop {}
  {\bibfield  {journal} {\bibinfo  {journal} {Phys. Rev. Lett.}\ }\textbf
  {\bibinfo {volume} {103}},\ \bibinfo {pages} {096405} (\bibinfo {year}
  {2009})}\BibitemShut {NoStop}%
\bibitem [{\citenamefont {Nolan}\ and\ \citenamefont
  {Elliott}(2008)}]{nolan_CM:2008}%
  \BibitemOpen
  \bibfield  {author} {\bibinfo {author} {\bibfnamefont {M.}~\bibnamefont
  {Nolan}}\ and\ \bibinfo {author} {\bibfnamefont {S.~D.}\ \bibnamefont
  {Elliott}},\ }\href@noop {} {\bibfield  {journal} {\bibinfo  {journal} {Chem.
  Mater.}\ }\textbf {\bibinfo {volume} {20}},\ \bibinfo {pages} {5522}
  (\bibinfo {year} {2008})}\BibitemShut {NoStop}%
\bibitem [{\citenamefont {Kawazoe}\ \emph {et~al.}(1997)\citenamefont
  {Kawazoe}, \citenamefont {Yasukawa}, \citenamefont {Hyodo}, \citenamefont
  {Kurita}, \citenamefont {Yanagi},\ and\ \citenamefont
  {Hosono}}]{kawazoe_Nature:1997}%
  \BibitemOpen
  \bibfield  {author} {\bibinfo {author} {\bibfnamefont {H.}~\bibnamefont
  {Kawazoe}}, \bibinfo {author} {\bibfnamefont {M.}~\bibnamefont {Yasukawa}},
  \bibinfo {author} {\bibfnamefont {H.}~\bibnamefont {Hyodo}}, \bibinfo
  {author} {\bibfnamefont {M.}~\bibnamefont {Kurita}}, \bibinfo {author}
  {\bibfnamefont {H.}~\bibnamefont {Yanagi}}, \ and\ \bibinfo {author}
  {\bibfnamefont {H.}~\bibnamefont {Hosono}},\ }\href@noop {} {\bibfield
  {journal} {\bibinfo  {journal} {Nature}\ }\textbf {\bibinfo {volume} {389}},\
  \bibinfo {pages} {939} (\bibinfo {year} {1997})}\BibitemShut {NoStop}%
\bibitem [{\citenamefont {Hautier}\ \emph {et~al.}(2013)\citenamefont
  {Hautier}, \citenamefont {Miglio}, \citenamefont {Ceder}, \citenamefont
  {Rignanese},\ and\ \citenamefont {Gonze}}]{hautier_NC:2013}%
  \BibitemOpen
  \bibfield  {author} {\bibinfo {author} {\bibfnamefont {G.}~\bibnamefont
  {Hautier}}, \bibinfo {author} {\bibfnamefont {A.}~\bibnamefont {Miglio}},
  \bibinfo {author} {\bibfnamefont {G.}~\bibnamefont {Ceder}}, \bibinfo
  {author} {\bibfnamefont {G.-M.}\ \bibnamefont {Rignanese}}, \ and\ \bibinfo
  {author} {\bibfnamefont {X.}~\bibnamefont {Gonze}},\ }\href@noop {}
  {\bibfield  {journal} {\bibinfo  {journal} {Nat. Commun.}\ }\textbf {\bibinfo
  {volume} {4}},\ \bibinfo {pages} {2292} (\bibinfo {year} {2013})}\BibitemShut
  {NoStop}%
\bibitem [{\citenamefont {Kale}\ \emph {et~al.}(2003)\citenamefont {Kale},
  \citenamefont {Ogale}, \citenamefont {Shinde}, \citenamefont {Sahasrabuddhe},
  \citenamefont {Kulkarni}, \citenamefont {Greene},\ and\ \citenamefont
  {Venkatesan}}]{kale_APL:2003}%
  \BibitemOpen
  \bibfield  {author} {\bibinfo {author} {\bibfnamefont {S.~N.}\ \bibnamefont
  {Kale}}, \bibinfo {author} {\bibfnamefont {S.~B.}\ \bibnamefont {Ogale}},
  \bibinfo {author} {\bibfnamefont {S.~R.}\ \bibnamefont {Shinde}}, \bibinfo
  {author} {\bibfnamefont {M.}~\bibnamefont {Sahasrabuddhe}}, \bibinfo {author}
  {\bibfnamefont {V.~N.}\ \bibnamefont {Kulkarni}}, \bibinfo {author}
  {\bibfnamefont {R.~L.}\ \bibnamefont {Greene}}, \ and\ \bibinfo {author}
  {\bibfnamefont {T.}~\bibnamefont {Venkatesan}},\ }\href@noop {} {\bibfield
  {journal} {\bibinfo  {journal} {Appl. Phys. Lett.}\ }\textbf {\bibinfo
  {volume} {82}},\ \bibinfo {pages} {2100} (\bibinfo {year}
  {2003})}\BibitemShut {NoStop}%
\bibitem [{\citenamefont {Raebiger}\ \emph
  {et~al.}(2007{\natexlab{b}})\citenamefont {Raebiger}, \citenamefont {Lany},\
  and\ \citenamefont {Zunger}}]{raebiger_PRL:2007}%
  \BibitemOpen
  \bibfield  {author} {\bibinfo {author} {\bibfnamefont {H.}~\bibnamefont
  {Raebiger}}, \bibinfo {author} {\bibfnamefont {S.}~\bibnamefont {Lany}}, \
  and\ \bibinfo {author} {\bibfnamefont {A.}~\bibnamefont {Zunger}},\
  }\href@noop {} {\bibfield  {journal} {\bibinfo  {journal} {Phys. Rev. Lett.}\
  }\textbf {\bibinfo {volume} {99}},\ \bibinfo {pages} {167203} (\bibinfo
  {year} {2007}{\natexlab{b}})}\BibitemShut {NoStop}%
\bibitem [{\citenamefont {Mittiga}\ \emph {et~al.}(2006)\citenamefont
  {Mittiga}, \citenamefont {Salza}, \citenamefont {Sarto}, \citenamefont
  {Tucci},\ and\ \citenamefont {Vasanthi}}]{mittiga_APL:2006}%
  \BibitemOpen
  \bibfield  {author} {\bibinfo {author} {\bibfnamefont {A.}~\bibnamefont
  {Mittiga}}, \bibinfo {author} {\bibfnamefont {E.}~\bibnamefont {Salza}},
  \bibinfo {author} {\bibfnamefont {F.}~\bibnamefont {Sarto}}, \bibinfo
  {author} {\bibfnamefont {M.}~\bibnamefont {Tucci}}, \ and\ \bibinfo {author}
  {\bibfnamefont {R.}~\bibnamefont {Vasanthi}},\ }\href@noop {} {\bibfield
  {journal} {\bibinfo  {journal} {Appl. Phys. Lett.}\ }\textbf {\bibinfo
  {volume} {88}},\ \bibinfo {pages} {163502} (\bibinfo {year}
  {2006})}\BibitemShut {NoStop}%
\bibitem [{\citenamefont {Meyer}\ \emph {et~al.}(2012)\citenamefont {Meyer},
  \citenamefont {Polity}, \citenamefont {Reppin}, \citenamefont {Becker},
  \citenamefont {Hering}, \citenamefont {Klar}, \citenamefont {Sander},
  \citenamefont {Reindl}, \citenamefont {Benz}, \citenamefont {Eickhoff},
  \citenamefont {Heiliger}, \citenamefont {Heinemann}, \citenamefont
  {Bl\"asing}, \citenamefont {Krost}, \citenamefont {Shokovets}, \citenamefont
  {M\"uller},\ and\ \citenamefont {Ronning}}]{meyer_PSS:2012}%
  \BibitemOpen
  \bibfield  {author} {\bibinfo {author} {\bibfnamefont {B.~K.}\ \bibnamefont
  {Meyer}}, \bibinfo {author} {\bibfnamefont {A.}~\bibnamefont {Polity}},
  \bibinfo {author} {\bibfnamefont {D.}~\bibnamefont {Reppin}}, \bibinfo
  {author} {\bibfnamefont {M.}~\bibnamefont {Becker}}, \bibinfo {author}
  {\bibfnamefont {P.}~\bibnamefont {Hering}}, \bibinfo {author} {\bibfnamefont
  {P.~J.}\ \bibnamefont {Klar}}, \bibinfo {author} {\bibfnamefont
  {T.}~\bibnamefont {Sander}}, \bibinfo {author} {\bibfnamefont
  {C.}~\bibnamefont {Reindl}}, \bibinfo {author} {\bibfnamefont
  {J.}~\bibnamefont {Benz}}, \bibinfo {author} {\bibfnamefont {M.}~\bibnamefont
  {Eickhoff}}, \bibinfo {author} {\bibfnamefont {C.}~\bibnamefont {Heiliger}},
  \bibinfo {author} {\bibfnamefont {M.}~\bibnamefont {Heinemann}}, \bibinfo
  {author} {\bibfnamefont {J.}~\bibnamefont {Bl\"asing}}, \bibinfo {author}
  {\bibfnamefont {A.}~\bibnamefont {Krost}}, \bibinfo {author} {\bibfnamefont
  {S.}~\bibnamefont {Shokovets}}, \bibinfo {author} {\bibfnamefont
  {C.}~\bibnamefont {M\"uller}}, \ and\ \bibinfo {author} {\bibfnamefont
  {C.}~\bibnamefont {Ronning}},\ }\href@noop {} {\bibfield  {journal} {\bibinfo
   {journal} {Phys. Status Solidi B}\ }\textbf {\bibinfo {volume} {249}},\
  \bibinfo {pages} {1487} (\bibinfo {year} {2012})}\BibitemShut {NoStop}%
\bibitem [{\citenamefont {Paracchino}\ \emph {et~al.}(2011)\citenamefont
  {Paracchino}, \citenamefont {Laporte}, \citenamefont {Sivula}, \citenamefont
  {Graetzel},\ and\ \citenamefont {Thimsen}}]{paracchino_NM:2011}%
  \BibitemOpen
  \bibfield  {author} {\bibinfo {author} {\bibfnamefont {A.}~\bibnamefont
  {Paracchino}}, \bibinfo {author} {\bibfnamefont {V.}~\bibnamefont {Laporte}},
  \bibinfo {author} {\bibfnamefont {K.}~\bibnamefont {Sivula}}, \bibinfo
  {author} {\bibfnamefont {M.}~\bibnamefont {Graetzel}}, \ and\ \bibinfo
  {author} {\bibfnamefont {E.}~\bibnamefont {Thimsen}},\ }\href@noop {}
  {\bibfield  {journal} {\bibinfo  {journal} {Nature Materials}\ }\textbf
  {\bibinfo {volume} {10}},\ \bibinfo {pages} {456} (\bibinfo {year}
  {2011})}\BibitemShut {NoStop}%
\bibitem [{\citenamefont {de~Walle}\ and\ \citenamefont
  {Neugebauer}(2004)}]{walle_JAP:2004}%
  \BibitemOpen
  \bibfield  {author} {\bibinfo {author} {\bibfnamefont {C.~G.~V.}\
  \bibnamefont {de~Walle}}\ and\ \bibinfo {author} {\bibfnamefont
  {J.}~\bibnamefont {Neugebauer}},\ }\href@noop {} {\bibfield  {journal}
  {\bibinfo  {journal} {J. Appl. Phys.}\ }\textbf {\bibinfo {volume} {95}},\
  \bibinfo {pages} {3851} (\bibinfo {year} {2004})}\BibitemShut {NoStop}%
\bibitem [{\citenamefont {Lany}\ and\ \citenamefont
  {Zunger}(2007)}]{lany_PRL:2007}%
  \BibitemOpen
  \bibfield  {author} {\bibinfo {author} {\bibfnamefont {S.}~\bibnamefont
  {Lany}}\ and\ \bibinfo {author} {\bibfnamefont {A.}~\bibnamefont {Zunger}},\
  }\href@noop {} {\bibfield  {journal} {\bibinfo  {journal} {Phys. Rev. Lett.}\
  }\textbf {\bibinfo {volume} {98}},\ \bibinfo {pages} {045501} (\bibinfo
  {year} {2007})}\BibitemShut {NoStop}%
\bibitem [{\citenamefont {Lany}\ and\ \citenamefont
  {Zunger}(2008)}]{lany_PRB:2008}%
  \BibitemOpen
  \bibfield  {author} {\bibinfo {author} {\bibfnamefont {S.}~\bibnamefont
  {Lany}}\ and\ \bibinfo {author} {\bibfnamefont {A.}~\bibnamefont {Zunger}},\
  }\href@noop {} {\bibfield  {journal} {\bibinfo  {journal} {Phys. Rev. B}\
  }\textbf {\bibinfo {volume} {78}},\ \bibinfo {pages} {235104} (\bibinfo
  {year} {2008})}\BibitemShut {NoStop}%
\bibitem [{\citenamefont {Agoston}\ \emph {et~al.}(2010)\citenamefont
  {Agoston}, \citenamefont {K\"{o}rber}, \citenamefont {Klein}, \citenamefont
  {Puska}, \citenamefont {Nieminen},\ and\ \citenamefont
  {Albe}}]{agoston_JAP:2010}%
  \BibitemOpen
  \bibfield  {author} {\bibinfo {author} {\bibfnamefont {P.}~\bibnamefont
  {Agoston}}, \bibinfo {author} {\bibfnamefont {C.}~\bibnamefont {K\"{o}rber}},
  \bibinfo {author} {\bibfnamefont {A.}~\bibnamefont {Klein}}, \bibinfo
  {author} {\bibfnamefont {M.~J.}\ \bibnamefont {Puska}}, \bibinfo {author}
  {\bibfnamefont {R.~M.}\ \bibnamefont {Nieminen}}, \ and\ \bibinfo {author}
  {\bibfnamefont {K.}~\bibnamefont {Albe}},\ }\href@noop {} {\bibfield
  {journal} {\bibinfo  {journal} {J. Appl. Phys.}\ }\textbf {\bibinfo {volume}
  {108}},\ \bibinfo {pages} {053511} (\bibinfo {year} {2010})}\BibitemShut
  {NoStop}%
\bibitem [{\citenamefont {Peng}\ \emph {et~al.}(2013)\citenamefont {Peng},
  \citenamefont {Scanlon}, \citenamefont {Stevanovic}, \citenamefont {Vidal},
  \citenamefont {Watson},\ and\ \citenamefont {Lany}}]{peng_PRB:2013}%
  \BibitemOpen
  \bibfield  {author} {\bibinfo {author} {\bibfnamefont {H.}~\bibnamefont
  {Peng}}, \bibinfo {author} {\bibfnamefont {D.~O.}\ \bibnamefont {Scanlon}},
  \bibinfo {author} {\bibfnamefont {V.}~\bibnamefont {Stevanovic}}, \bibinfo
  {author} {\bibfnamefont {J.}~\bibnamefont {Vidal}}, \bibinfo {author}
  {\bibfnamefont {G.~W.}\ \bibnamefont {Watson}}, \ and\ \bibinfo {author}
  {\bibfnamefont {S.}~\bibnamefont {Lany}},\ }\href@noop {} {\bibfield
  {journal} {\bibinfo  {journal} {Phys. Rev. B}\ }\textbf {\bibinfo {volume}
  {88}},\ \bibinfo {pages} {115201} (\bibinfo {year} {2013})}\BibitemShut
  {NoStop}%
\bibitem [{\citenamefont {Biswas}\ and\ \citenamefont
  {Lany}(2009)}]{biswas_PRB:2009}%
  \BibitemOpen
  \bibfield  {author} {\bibinfo {author} {\bibfnamefont {K.}~\bibnamefont
  {Biswas}}\ and\ \bibinfo {author} {\bibfnamefont {S.}~\bibnamefont {Lany}},\
  }\href@noop {} {\bibfield  {journal} {\bibinfo  {journal} {Phys. Rev. B}\
  }\textbf {\bibinfo {volume} {80}},\ \bibinfo {pages} {115206} (\bibinfo
  {year} {2009})}\BibitemShut {NoStop}%
\bibitem [{\citenamefont {Wright}\ and\ \citenamefont
  {Nelson}(2002)}]{wright_JAP:2002}%
  \BibitemOpen
  \bibfield  {author} {\bibinfo {author} {\bibfnamefont {A.~F.}\ \bibnamefont
  {Wright}}\ and\ \bibinfo {author} {\bibfnamefont {J.~S.}\ \bibnamefont
  {Nelson}},\ }\href@noop {} {\bibfield  {journal} {\bibinfo  {journal} {J.
  Appl. Phys.}\ }\textbf {\bibinfo {volume} {92}},\ \bibinfo {pages} {5849}
  (\bibinfo {year} {2002})}\BibitemShut {NoStop}%
\bibitem [{\citenamefont {Perkins}\ \emph {et~al.}(1999)\citenamefont
  {Perkins}, \citenamefont {Mascarenhas}, \citenamefont {Zhang}, \citenamefont
  {Geisz}, \citenamefont {Friedman}, \citenamefont {Olson},\ and\ \citenamefont
  {Kurtz}}]{perkins_PRL:1999}%
  \BibitemOpen
  \bibfield  {author} {\bibinfo {author} {\bibfnamefont {J.~D.}\ \bibnamefont
  {Perkins}}, \bibinfo {author} {\bibfnamefont {A.}~\bibnamefont
  {Mascarenhas}}, \bibinfo {author} {\bibfnamefont {Y.}~\bibnamefont {Zhang}},
  \bibinfo {author} {\bibfnamefont {J.}~\bibnamefont {Geisz}}, \bibinfo
  {author} {\bibfnamefont {D.}~\bibnamefont {Friedman}}, \bibinfo {author}
  {\bibfnamefont {J.}~\bibnamefont {Olson}}, \ and\ \bibinfo {author}
  {\bibfnamefont {S.}~\bibnamefont {Kurtz}},\ }\href@noop {} {\bibfield
  {journal} {\bibinfo  {journal} {Phys. Rev. Lett.}\ }\textbf {\bibinfo
  {volume} {82}},\ \bibinfo {pages} {3312} (\bibinfo {year}
  {1999})}\BibitemShut {NoStop}%
\bibitem [{\citenamefont {Lany}\ \emph {et~al.}(2005)\citenamefont {Lany},
  \citenamefont {Zhao}, \citenamefont {Persson},\ and\ \citenamefont
  {Zunger}}]{lany_APL:2005}%
  \BibitemOpen
  \bibfield  {author} {\bibinfo {author} {\bibfnamefont {S.}~\bibnamefont
  {Lany}}, \bibinfo {author} {\bibfnamefont {Y.-J.}\ \bibnamefont {Zhao}},
  \bibinfo {author} {\bibfnamefont {C.}~\bibnamefont {Persson}}, \ and\
  \bibinfo {author} {\bibfnamefont {A.}~\bibnamefont {Zunger}},\ }\href@noop {}
  {\bibfield  {journal} {\bibinfo  {journal} {Appl. Phys. Lett.}\ }\textbf
  {\bibinfo {volume} {86}},\ \bibinfo {pages} {042109} (\bibinfo {year}
  {2005})}\BibitemShut {NoStop}%
\bibitem [{\citenamefont {Mittiga}\ \emph {et~al.}(2009)\citenamefont
  {Mittiga}, \citenamefont {Biccari},\ and\ \citenamefont
  {Malerba}}]{mittiga_TSF:2009}%
  \BibitemOpen
  \bibfield  {author} {\bibinfo {author} {\bibfnamefont {A.}~\bibnamefont
  {Mittiga}}, \bibinfo {author} {\bibfnamefont {F.}~\bibnamefont {Biccari}}, \
  and\ \bibinfo {author} {\bibfnamefont {C.}~\bibnamefont {Malerba}},\
  }\href@noop {} {\bibfield  {journal} {\bibinfo  {journal} {Thin Solid Films}\
  }\textbf {\bibinfo {volume} {517}},\ \bibinfo {pages} {2469} (\bibinfo {year}
  {2009})}\BibitemShut {NoStop}%
\bibitem [{\citenamefont {Desnica}(1998)}]{Desnica_98}%
  \BibitemOpen
  \bibfield  {author} {\bibinfo {author} {\bibfnamefont {U.~V.}\ \bibnamefont
  {Desnica}},\ }\href@noop {} {\bibfield  {journal} {\bibinfo  {journal}
  {Progress in Crystal Growth and Characterization of Materials}\ }\textbf
  {\bibinfo {volume} {36}},\ \bibinfo {pages} {291} (\bibinfo {year}
  {1998})}\BibitemShut {NoStop}%
\bibitem [{\citenamefont {Zakutayev}\ \emph {et~al.}(2012)\citenamefont
  {Zakutayev}, \citenamefont {Paudel}, \citenamefont {Ndione}, \citenamefont
  {Perkins}, \citenamefont {Lany}, \citenamefont {Zunger},\ and\ \citenamefont
  {Ginley}}]{zak_PRB:2012}%
  \BibitemOpen
  \bibfield  {author} {\bibinfo {author} {\bibfnamefont {A.}~\bibnamefont
  {Zakutayev}}, \bibinfo {author} {\bibfnamefont {T.~R.}\ \bibnamefont
  {Paudel}}, \bibinfo {author} {\bibfnamefont {P.~F.}\ \bibnamefont {Ndione}},
  \bibinfo {author} {\bibfnamefont {J.~D.}\ \bibnamefont {Perkins}}, \bibinfo
  {author} {\bibfnamefont {S.}~\bibnamefont {Lany}}, \bibinfo {author}
  {\bibfnamefont {A.}~\bibnamefont {Zunger}}, \ and\ \bibinfo {author}
  {\bibfnamefont {D.}~\bibnamefont {Ginley}},\ }\href@noop {} {\bibfield
  {journal} {\bibinfo  {journal} {Phys. Rev. B}\ }\textbf {\bibinfo {volume}
  {85}},\ \bibinfo {pages} {085204} (\bibinfo {year} {2012})}\BibitemShut
  {NoStop}%
\bibitem [{\citenamefont {Gonz\'alez}\ \emph {et~al.}(2012)\citenamefont
  {Gonz\'alez}, \citenamefont {Mason}, \citenamefont {Okasinski}, \citenamefont
  {Buslaps},\ and\ \citenamefont {Honkim\"aki}}]{Mason_12}%
  \BibitemOpen
  \bibfield  {author} {\bibinfo {author} {\bibfnamefont {G.~B.}\ \bibnamefont
  {Gonz\'alez}}, \bibinfo {author} {\bibfnamefont {T.~O.}\ \bibnamefont
  {Mason}}, \bibinfo {author} {\bibfnamefont {J.~S.}\ \bibnamefont
  {Okasinski}}, \bibinfo {author} {\bibfnamefont {T.}~\bibnamefont {Buslaps}},
  \ and\ \bibinfo {author} {\bibfnamefont {V.}~\bibnamefont {Honkim\"aki}},\
  }\href@noop {} {\bibfield  {journal} {\bibinfo  {journal} {J. Am. Ceram.
  Soc.}\ }\textbf {\bibinfo {volume} {95}},\ \bibinfo {pages} {809} (\bibinfo
  {year} {2012})}\BibitemShut {NoStop}%
\bibitem [{\citenamefont {Singh}\ \emph {et~al.}(1997)\citenamefont {Singh},
  \citenamefont {Doppalapudi}, \citenamefont {Moustakas},\ and\ \citenamefont
  {Romano}}]{singh_APL:1997}%
  \BibitemOpen
  \bibfield  {author} {\bibinfo {author} {\bibfnamefont {R.}~\bibnamefont
  {Singh}}, \bibinfo {author} {\bibfnamefont {D.}~\bibnamefont {Doppalapudi}},
  \bibinfo {author} {\bibfnamefont {T.~D.}\ \bibnamefont {Moustakas}}, \ and\
  \bibinfo {author} {\bibfnamefont {L.~T.}\ \bibnamefont {Romano}},\
  }\href@noop {} {\bibfield  {journal} {\bibinfo  {journal} {Appl. Phys.
  Lett.}\ }\textbf {\bibinfo {volume} {70}},\ \bibinfo {pages} {1089} (\bibinfo
  {year} {1997})}\BibitemShut {NoStop}%
\bibitem [{\citenamefont {Lany}\ and\ \citenamefont
  {Zunger}(2009)}]{lany_MSSE:2009}%
  \BibitemOpen
  \bibfield  {author} {\bibinfo {author} {\bibfnamefont {S.}~\bibnamefont
  {Lany}}\ and\ \bibinfo {author} {\bibfnamefont {A.}~\bibnamefont {Zunger}},\
  }\href@noop {} {\bibfield  {journal} {\bibinfo  {journal} {Modelling Simul.
  Sci. Eng.}\ }\textbf {\bibinfo {volume} {17}},\ \bibinfo {pages} {084002}
  (\bibinfo {year} {2009})}\BibitemShut {NoStop}%
\bibitem [{\citenamefont {Zakutayev}\ \emph
  {et~al.}(2013{\natexlab{a}})\citenamefont {Zakutayev}, \citenamefont {Perry},
  \citenamefont {Mason}, \citenamefont {Ginley},\ and\ \citenamefont
  {Lany}}]{zakutayev_APL:2013}%
  \BibitemOpen
  \bibfield  {author} {\bibinfo {author} {\bibfnamefont {A.}~\bibnamefont
  {Zakutayev}}, \bibinfo {author} {\bibfnamefont {N.~H.}\ \bibnamefont
  {Perry}}, \bibinfo {author} {\bibfnamefont {T.~O.}\ \bibnamefont {Mason}},
  \bibinfo {author} {\bibfnamefont {D.~S.}\ \bibnamefont {Ginley}}, \ and\
  \bibinfo {author} {\bibfnamefont {S.}~\bibnamefont {Lany}},\ }\href@noop {}
  {\bibfield  {journal} {\bibinfo  {journal} {Appl. Phys. Lett.}\ }\textbf
  {\bibinfo {volume} {103}},\ \bibinfo {pages} {232106} (\bibinfo {year}
  {2013}{\natexlab{a}})}\BibitemShut {NoStop}%
\bibitem [{\citenamefont {Zakutayev}\ \emph
  {et~al.}(2013{\natexlab{b}})\citenamefont {Zakutayev}, \citenamefont
  {Luciano}, \citenamefont {Bollinger}, \citenamefont {Sigdel}, \citenamefont
  {Ndione}, \citenamefont {Perkins}, \citenamefont {Berry}, \citenamefont
  {Parilla},\ and\ \citenamefont {Ginley}}]{zak_RSI:2013}%
  \BibitemOpen
  \bibfield  {author} {\bibinfo {author} {\bibfnamefont {A.}~\bibnamefont
  {Zakutayev}}, \bibinfo {author} {\bibfnamefont {F.~J.}\ \bibnamefont
  {Luciano}}, \bibinfo {author} {\bibfnamefont {V.~P.}\ \bibnamefont
  {Bollinger}}, \bibinfo {author} {\bibfnamefont {A.~K.}\ \bibnamefont
  {Sigdel}}, \bibinfo {author} {\bibfnamefont {P.~F.}\ \bibnamefont {Ndione}},
  \bibinfo {author} {\bibfnamefont {J.~D.}\ \bibnamefont {Perkins}}, \bibinfo
  {author} {\bibfnamefont {J.~J.}\ \bibnamefont {Berry}}, \bibinfo {author}
  {\bibfnamefont {P.~A.}\ \bibnamefont {Parilla}}, \ and\ \bibinfo {author}
  {\bibfnamefont {D.~S.}\ \bibnamefont {Ginley}},\ }\href@noop {} {\bibfield
  {journal} {\bibinfo  {journal} {Rev. Sci. Instr.}\ }\textbf {\bibinfo
  {volume} {84}},\ \bibinfo {pages} {053905} (\bibinfo {year}
  {2013}{\natexlab{b}})}\BibitemShut {NoStop}%
\bibitem [{\citenamefont {Zakutayev}\ \emph {et~al.}(2011)\citenamefont
  {Zakutayev}, \citenamefont {Perkins}, \citenamefont {Parilla}, \citenamefont
  {Widjonarko}, \citenamefont {Sigdel}, \citenamefont {Berry},\ and\
  \citenamefont {Ginley}}]{zak_MRSC:2011}%
  \BibitemOpen
  \bibfield  {author} {\bibinfo {author} {\bibfnamefont {A.}~\bibnamefont
  {Zakutayev}}, \bibinfo {author} {\bibfnamefont {J.~D.}\ \bibnamefont
  {Perkins}}, \bibinfo {author} {\bibfnamefont {P.~A.}\ \bibnamefont
  {Parilla}}, \bibinfo {author} {\bibfnamefont {N.~E.}\ \bibnamefont
  {Widjonarko}}, \bibinfo {author} {\bibfnamefont {A.~K.}\ \bibnamefont
  {Sigdel}}, \bibinfo {author} {\bibfnamefont {J.~J.}\ \bibnamefont {Berry}}, \
  and\ \bibinfo {author} {\bibfnamefont {D.}~\bibnamefont {Ginley}},\
  }\href@noop {} {\bibfield  {journal} {\bibinfo  {journal} {MRS
  Communications}\ }\textbf {\bibinfo {volume} {1}},\ \bibinfo {pages} {23}
  (\bibinfo {year} {2011})}\BibitemShut {NoStop}%
\bibitem [{\citenamefont {Mandel}(1964)}]{mandel_PR:1964}%
  \BibitemOpen
  \bibfield  {author} {\bibinfo {author} {\bibfnamefont {G.}~\bibnamefont
  {Mandel}},\ }\href@noop {} {\bibfield  {journal} {\bibinfo  {journal} {Phys.
  Rev.}\ }\textbf {\bibinfo {volume} {134}},\ \bibinfo {pages} {A1073}
  (\bibinfo {year} {1964})}\BibitemShut {NoStop}%
\bibitem [{\citenamefont {Malerba}\ \emph {et~al.}(2011)\citenamefont
  {Malerba}, \citenamefont {Biccari}, \citenamefont {Ricardo}, \citenamefont
  {DÕIncau}, \citenamefont {Scardi},\ and\ \citenamefont
  {Mittiga}}]{malerba_SEMSC:2011}%
  \BibitemOpen
  \bibfield  {author} {\bibinfo {author} {\bibfnamefont {C.}~\bibnamefont
  {Malerba}}, \bibinfo {author} {\bibfnamefont {F.}~\bibnamefont {Biccari}},
  \bibinfo {author} {\bibfnamefont {C.~L.~A.}\ \bibnamefont {Ricardo}},
  \bibinfo {author} {\bibfnamefont {M.}~\bibnamefont {DÕIncau}}, \bibinfo
  {author} {\bibfnamefont {P.}~\bibnamefont {Scardi}}, \ and\ \bibinfo {author}
  {\bibfnamefont {A.}~\bibnamefont {Mittiga}},\ }\href@noop {} {\bibfield
  {journal} {\bibinfo  {journal} {Sol. En. Mater. Sol. Cells}\ }\textbf
  {\bibinfo {volume} {95}},\ \bibinfo {pages} {2848} (\bibinfo {year}
  {2011})}\BibitemShut {NoStop}%
\bibitem [{\citenamefont {Papadimitriou}(1993)}]{papadimitriou_SSE:1993}%
  \BibitemOpen
  \bibfield  {author} {\bibinfo {author} {\bibfnamefont {L.}~\bibnamefont
  {Papadimitriou}},\ }\href@noop {} {\bibfield  {journal} {\bibinfo  {journal}
  {Solid-State Electron.}\ }\textbf {\bibinfo {volume} {36}},\ \bibinfo {pages}
  {431} (\bibinfo {year} {1993})}\BibitemShut {NoStop}%
\bibitem [{\citenamefont {Bl\"ochl}(1994)}]{blochl_PRB:1994}%
  \BibitemOpen
  \bibfield  {author} {\bibinfo {author} {\bibfnamefont {P.~E.}\ \bibnamefont
  {Bl\"ochl}},\ }\href@noop {} {\bibfield  {journal} {\bibinfo  {journal}
  {Phys. Rev. B}\ }\textbf {\bibinfo {volume} {50}},\ \bibinfo {pages} {17953}
  (\bibinfo {year} {1994})}\BibitemShut {NoStop}%
\bibitem [{\citenamefont {Kresse}\ and\ \citenamefont
  {Joubert}(1999)}]{joubert_PRB:1999}%
  \BibitemOpen
  \bibfield  {author} {\bibinfo {author} {\bibfnamefont {G.}~\bibnamefont
  {Kresse}}\ and\ \bibinfo {author} {\bibfnamefont {D.}~\bibnamefont
  {Joubert}},\ }\href@noop {} {\bibfield  {journal} {\bibinfo  {journal} {Phys.
  Rev. B}\ }\textbf {\bibinfo {volume} {59}},\ \bibinfo {pages} {1758}
  (\bibinfo {year} {1999})}\BibitemShut {NoStop}%
\bibitem [{\citenamefont {Perdew}\ \emph {et~al.}(1996)\citenamefont {Perdew},
  \citenamefont {Burke},\ and\ \citenamefont {Ernzerhof}}]{perdew_PRL:1996}%
  \BibitemOpen
  \bibfield  {author} {\bibinfo {author} {\bibfnamefont {J.~P.}\ \bibnamefont
  {Perdew}}, \bibinfo {author} {\bibfnamefont {K.}~\bibnamefont {Burke}}, \
  and\ \bibinfo {author} {\bibfnamefont {M.}~\bibnamefont {Ernzerhof}},\
  }\href@noop {} {\bibfield  {journal} {\bibinfo  {journal} {Phys. Rev. Lett.}\
  }\textbf {\bibinfo {volume} {77}},\ \bibinfo {pages} {3865} (\bibinfo {year}
  {1996})}\BibitemShut {NoStop}%
\bibitem [{\citenamefont {Dudarev}\ \emph {et~al.}(1998)\citenamefont
  {Dudarev}, \citenamefont {Botton}, \citenamefont {Savrasov}, \citenamefont
  {Humphreys},\ and\ \citenamefont {Sutton}}]{dudarev_PRB:1998}%
  \BibitemOpen
  \bibfield  {author} {\bibinfo {author} {\bibfnamefont {S.~L.}\ \bibnamefont
  {Dudarev}}, \bibinfo {author} {\bibfnamefont {G.~A.}\ \bibnamefont {Botton}},
  \bibinfo {author} {\bibfnamefont {S.~Y.}\ \bibnamefont {Savrasov}}, \bibinfo
  {author} {\bibfnamefont {C.~J.}\ \bibnamefont {Humphreys}}, \ and\ \bibinfo
  {author} {\bibfnamefont {A.~P.}\ \bibnamefont {Sutton}},\ }\href@noop {}
  {\bibfield  {journal} {\bibinfo  {journal} {Phys. Rev. B}\ }\textbf {\bibinfo
  {volume} {57}},\ \bibinfo {pages} {1505} (\bibinfo {year}
  {1998})}\BibitemShut {NoStop}%
\bibitem [{\citenamefont {Lany}(2013)}]{lany_3dGW:2013}%
  \BibitemOpen
  \bibfield  {author} {\bibinfo {author} {\bibfnamefont {S.}~\bibnamefont
  {Lany}},\ }\href@noop {} {\bibfield  {journal} {\bibinfo  {journal} {Phys.
  Rev. B}\ }\textbf {\bibinfo {volume} {87}},\ \bibinfo {pages} {085112}
  (\bibinfo {year} {2013})}\BibitemShut {NoStop}%
\bibitem [{\citenamefont {Stevanovi\'c}\ \emph {et~al.}(2012)\citenamefont
  {Stevanovi\'c}, \citenamefont {Lany}, \citenamefont {Zhang},\ and\
  \citenamefont {Zunger}}]{vladan_FERE}%
  \BibitemOpen
  \bibfield  {author} {\bibinfo {author} {\bibfnamefont {V.}~\bibnamefont
  {Stevanovi\'c}}, \bibinfo {author} {\bibfnamefont {S.}~\bibnamefont {Lany}},
  \bibinfo {author} {\bibfnamefont {X.}~\bibnamefont {Zhang}}, \ and\ \bibinfo
  {author} {\bibfnamefont {A.}~\bibnamefont {Zunger}},\ }\href@noop {}
  {\bibfield  {journal} {\bibinfo  {journal} {Phys. Rev. B}\ }\textbf {\bibinfo
  {volume} {85}},\ \bibinfo {pages} {115104} (\bibinfo {year}
  {2012})}\BibitemShut {NoStop}%
\bibitem [{\citenamefont {Shishkin}\ and\ \citenamefont
  {Kresse}(2006)}]{shishkin_gwPRB:2006}%
  \BibitemOpen
  \bibfield  {author} {\bibinfo {author} {\bibfnamefont {M.}~\bibnamefont
  {Shishkin}}\ and\ \bibinfo {author} {\bibfnamefont {G.}~\bibnamefont
  {Kresse}},\ }\href@noop {} {\bibfield  {journal} {\bibinfo  {journal} {Phys.
  Rev. B}\ }\textbf {\bibinfo {volume} {74}},\ \bibinfo {pages} {035101}
  (\bibinfo {year} {2006})}\BibitemShut {NoStop}%
\bibitem [{\citenamefont {Samsonidze}\ \emph {et~al.}(2011)\citenamefont
  {Samsonidze}, \citenamefont {Jain}, \citenamefont {Deslippe}, \citenamefont
  {Cohen},\ and\ \citenamefont {Louie}}]{samsonidze_PRL:2011}%
  \BibitemOpen
  \bibfield  {author} {\bibinfo {author} {\bibfnamefont {G.}~\bibnamefont
  {Samsonidze}}, \bibinfo {author} {\bibfnamefont {M.}~\bibnamefont {Jain}},
  \bibinfo {author} {\bibfnamefont {J.}~\bibnamefont {Deslippe}}, \bibinfo
  {author} {\bibfnamefont {M.~L.}\ \bibnamefont {Cohen}}, \ and\ \bibinfo
  {author} {\bibfnamefont {S.~G.}\ \bibnamefont {Louie}},\ }\href@noop {}
  {\bibfield  {journal} {\bibinfo  {journal} {Phys. Rev. Lett.}\ }\textbf
  {\bibinfo {volume} {107}},\ \bibinfo {pages} {186404} (\bibinfo {year}
  {2011})}\BibitemShut {NoStop}%
\bibitem [{\citenamefont {Friedrich}\ \emph {et~al.}(2011)\citenamefont
  {Friedrich}, \citenamefont {M\"uller},\ and\ \citenamefont
  {Bl\"ugel}}]{friedrich_PRB:2011}%
  \BibitemOpen
  \bibfield  {author} {\bibinfo {author} {\bibfnamefont {C.}~\bibnamefont
  {Friedrich}}, \bibinfo {author} {\bibfnamefont {M.~C.}\ \bibnamefont
  {M\"uller}}, \ and\ \bibinfo {author} {\bibfnamefont {S.}~\bibnamefont
  {Bl\"ugel}},\ }\href@noop {} {\bibfield  {journal} {\bibinfo  {journal}
  {Phys. Rev. B}\ }\textbf {\bibinfo {volume} {83}},\ \bibinfo {pages}
  {081101(R)} (\bibinfo {year} {2011})}\BibitemShut {NoStop}%
\bibitem [{\citenamefont {Klime\v{s}}\ \emph {et~al.}()\citenamefont
  {Klime\v{s}}, \citenamefont {Kaltak},\ and\ \citenamefont
  {Kresse}}]{klimes_arx:2014}%
  \BibitemOpen
  \bibfield  {author} {\bibinfo {author} {\bibfnamefont {J.}~\bibnamefont
  {Klime\v{s}}}, \bibinfo {author} {\bibfnamefont {M.}~\bibnamefont {Kaltak}},
  \ and\ \bibinfo {author} {\bibfnamefont {G.}~\bibnamefont {Kresse}},\
  }\href@noop {} {\bibinfo  {journal} {arXiv:1404.3101}\ }\BibitemShut
  {NoStop}%
\bibitem [{\citenamefont {Stevanovi\'c}\ \emph {et~al.}(2014)\citenamefont
  {Stevanovi\'c}, \citenamefont {Lany}, \citenamefont {Ginley}, \citenamefont
  {Tumas},\ and\ \citenamefont {Zunger}}]{stevanovic_PCCP:2013}%
  \BibitemOpen
\bibfield  {journal} {  }\bibfield  {author} {\bibinfo {author} {\bibfnamefont
  {V.}~\bibnamefont {Stevanovi\'c}}, \bibinfo {author} {\bibfnamefont
  {S.}~\bibnamefont {Lany}}, \bibinfo {author} {\bibfnamefont {D.~S.}\
  \bibnamefont {Ginley}}, \bibinfo {author} {\bibfnamefont {W.}~\bibnamefont
  {Tumas}}, \ and\ \bibinfo {author} {\bibfnamefont {A.}~\bibnamefont
  {Zunger}},\ }\href@noop {} {\bibfield  {journal} {\bibinfo  {journal} {Phys.
  Chem. Chem. Phys.}\ }\textbf {\bibinfo {volume} {16}},\ \bibinfo {pages}
  {3706} (\bibinfo {year} {2014})}\BibitemShut {NoStop}%
\bibitem [{\citenamefont {Iwamitsu}\ \emph {et~al.}(2012)\citenamefont
  {Iwamitsu}, \citenamefont {Aihara}, \citenamefont {Shimamoto}, \citenamefont
  {Fujii},\ and\ \citenamefont {Akai}}]{iwamitsu_PSSC:2012}%
  \BibitemOpen
  \bibfield  {author} {\bibinfo {author} {\bibfnamefont {K.}~\bibnamefont
  {Iwamitsu}}, \bibinfo {author} {\bibfnamefont {S.}~\bibnamefont {Aihara}},
  \bibinfo {author} {\bibfnamefont {T.}~\bibnamefont {Shimamoto}}, \bibinfo
  {author} {\bibfnamefont {A.}~\bibnamefont {Fujii}}, \ and\ \bibinfo {author}
  {\bibfnamefont {I.}~\bibnamefont {Akai}},\ }\href@noop {} {\bibfield
  {journal} {\bibinfo  {journal} {physica status solidi (c)}\ }\textbf
  {\bibinfo {volume} {9}},\ \bibinfo {pages} {1610} (\bibinfo {year}
  {2012})}\BibitemShut {NoStop}%
\end{thebibliography}
%
\end{document}